\documentclass[journal]{IEEEtran}
\usepackage{amsmath,amsfonts}
\usepackage{bm}
\usepackage{algorithmic}
\usepackage{array}
\usepackage[caption=false,font=normalsize,labelfont=sf,textfont=sf]{subfig}
\usepackage{textcomp}
\usepackage{stfloats}
\usepackage{url}
\usepackage{verbatim}
\usepackage{graphicx}
\usepackage{multirow}
\usepackage{booktabs}
\usepackage{xfakebold}
\usepackage{color,soul}
\soulregister\cite7
\soulregister\ref7
\soulregister\eqref7
\usepackage{algorithmic}
\usepackage{tcolorbox}
\newcommand{\setBoldness}[1]{\def\fake@bold{#1}}
\newcommand{\fbseries}{\unskip\setBold\aftergroup\unsetBold\aftergroup\ignorespaces}

\newcommand{\myBoldness}{0.5}
\newcommand{\boldE}{\setBold[\myBoldness] \textit{e} \unsetBold}

\newcommand{\boldM}{\setBold[\myBoldness] \textit{m} \unsetBold}
\newcommand{\boldH}{\setBold[\myBoldness] \textit{h} \unsetBold}

\newcommand{\boldW}{\setBold[\myBoldness] \textit{w} \unsetBold}

\usepackage[flushleft]{threeparttable}

\usepackage[table]{xcolor}
\usepackage[ruled,vlined]{algorithm2e}

\begin{document}

% \title{Trained Reward Summarizer: Unsupervised Video Summarization through Reinforcement Learning with a Learned Reward Function}
% \title{Unsupervised Video Summarization via Reinforcement Learning Using a Trained Reward Generator}
\title{Reinforcement Learning for Unsupervised Video Summarization with Reward Generator Training}
% \title{Harnessing a Trained Reward Generator for Reinforcement Learning in Unsupervised Video Summarization}
% \title{Reinforcement Learning and Trained Reward Generator: A New Paradigm in Unsupervised Video Summarization}
% \title{Unsupervised Video Summarization: A Reinforcement Learning Approach with a Trained Evaluator}
% \title{Leveraging a Trained Evaluator for Reinforcement Learning in Unsupervised Video Summarization}
% \title{Learnable Summary Evaluator: A New Approach to Unsupervised Video Summarization via Reinforcement Learning}

\author{Mehryar Abbasi,~\IEEEmembership{Student Member,~IEEE,} Hadi Hadizadeh~\IEEEmembership{ Member,~IEEE}, Parvaneh Saeedi,~\IEEEmembership{ Member,~IEEE}
}
% \author{...}
        % <-this % stops a space

\IEEEpubid{\begin{tabular}[t]{@{}l@{}} \copyright~2025 IEEE. Personal use of this material is permitted. However, permission to use this material for \\ any other purposes must be obtained from the IEEE by sending an email to pubs-permissions@ieee.org. \end{tabular}}

\maketitle

\begin{abstract}

This paper presents a novel approach for unsupervised video summarization using reinforcement learning (RL), addressing limitations like unstable adversarial training and reliance on heuristic-based reward functions. The method operates on the principle that reconstruction fidelity serves as a proxy for informativeness, correlating summary quality with reconstruction ability. The summarizer model assigns importance scores to frames to generate the final summary. For training, RL is coupled with a unique reward generation pipeline that incentivizes improved reconstructions. This pipeline uses a generator model to reconstruct the full video from the selected summary frames; the similarity between the original and reconstructed video provides the reward signal. The generator itself is pre-trained self-supervisedly to reconstruct randomly masked frames. This two-stage training process enhances stability compared to adversarial architectures. Experimental results show strong alignment with human judgments and promising F-scores, validating the reconstruction objective. {The code for this project will be available online\footnote{\url{https://github.com/mehryar72/TR-SUM}}}.\let\thefootnote\relax\footnotetext{Mehryar Abbasi, Hadi Hadizadeh, and Parvaneh Saeedi are with School Of Engineering Science, Simon Fraser University, Canada (e-mail: \texttt{mabbasib@sfu.ca}; \texttt{hadi\_hadizadeh@sfu.ca}; \texttt{psaeedi@sfu.ca}).}

\end{abstract}

\begin{IEEEkeywords}
Unsupervised Video summarization, self-supervised learning, Reinforcement Learning, Transformers.
\end{IEEEkeywords}

\section{Introduction}\label{sec:int}

\IEEEPARstart{V}{ideo} summarization provides a condensed representation of video content and enables users to grasp its core essence swiftly. With the surge of video data, the demand for more efficient methods for indexing, searching, and managing extensive video databases becomes increasingly urgent~\cite{deepai, frontiers}. Video summarization provides condensed content in surveillance systems, online learning platforms, and social media~\cite{gygli2016}. It aids in identifying events in traffic monitoring systems~\cite{trafic}, serves as a resource in healthcare and education~\cite{surg, gupta2023}, and assists us in navigating through the immense volume of video data~\cite{springer}. 

{Videos can be summarized in two main ways: key frame selection to create a static storyboard of significant moments, or short segment compilation to produce a dynamic video skim showcasing key scenes~\cite{jadon2020unsupervised}}. A common guideline for this process is that the summary length should not exceed 15\% of the input video length~\cite{surv}, ensuring that it captures the most critical aspects while remaining concise and easy to watch. {In recent years, deep learning-based methods for automated video summarization have gained popularity~\cite{surv}, often outperforming conventional approaches (e.g.,~\cite{ovp,ma2020similarity}) \IEEEpubidadjcol that rely on sparse subset selection or clustering algorithms.} However, many of these methods rely on human-generated labels to train their models~\cite{ghauri2021supervised,zhu2022relational,apostolidis2021combining}, leading to challenges with scarcity, subjectivity, and bias in human annotations. As a result, there has been a focus on developing unsupervised video summarization methods~\cite{mahasseni2017unsupervised, apostolidis2019stepwise,yuan2019cycle, jung2019discriminative,apostolidis2020ac, apostolidis2020unsupervised,jung2020global,liu2019learning,he2019unsupervised, zhou2018deep, gonuguntla2019enhanced, zhao2019property,yoon2021interp,SummarizingACheat,yaliniz2021using,myicip2023sum}.
Unsupervised methods do not require human annotations. Instead, they use heuristic criteria such as diversity, coverage, and/or representativeness to select summary frames. However, these methods often fail to capture the semantic relevance and coherence of the summary and may produce redundant or noisy frames. Some of the existing works use complex or unstable architectures (e.g., RNNs, LSTMs, GANs) and training procedures (e.g. adversarial learning)
~\cite{mahasseni2017unsupervised, apostolidis2019stepwise,yuan2019cycle, jung2019discriminative,apostolidis2020ac, apostolidis2020unsupervised,jung2020global,liu2019learning,he2019unsupervised}. Other methods employ training criteria, reward, and loss functions that do not strongly correlate with the way a human would generate a video summary, thereby limiting their performance metrics~\cite{zhou2018deep, gonuguntla2019enhanced, zhao2019property,yoon2021interp,SummarizingACheat}.

{To address the limitations of prior work, we propose TR-SUM, an unsupervised video summarization framework that learns to assign frame importance scores using reinforcement learning, guided by a novel reward generation pipeline. Unlike earlier methods that rely on RNNs or GANs with unstable training dynamics, TR-SUM employs a transformer-based summarizer to model long-range temporal dependencies better. It avoids adversarial training by adopting a two-stage strategy: (1) a self-supervised generator is trained to reconstruct randomly masked video segments using a dynamic window masking strategy; (2) a summarizer is then trained using reinforcement learning, guided by reconstruction-based reward signals. During this stage, frame scores are interpreted as Bernoulli sampling probabilities to generate masked inputs for the generator. The reconstruction loss is converted into a reward, encouraging the summarizer to prioritize frames that enhance the reconstruction quality. This generator-guided reward offers a semantically grounded signal that aligns better with human summary expectations than hand-crafted objectives.

% This work extends our previous method, RS-SUM~\cite{myicip2023sum}, which used a self-supervised generator to infer frame importance by iteratively masking each frame. While effective, RS-SUM lacked a dedicated summarizer and suffered from high inference cost. TR-SUM addresses these limitations with a more efficient architecture, a dedicated single-pass scoring model and a new training pipeline. The generator is used exclusively in the reward generation stage, and its training is upgraded with a dynamic window masking strategy.

Based on your request to improve the writing while preserving the original meaning, here is a revised version of the paragraph. The edits focus on correcting typos, improving sentence structure, and enhancing clarity.

% {The idea of linking reconstruction quality to information representativeness finds its roots in foundational video summarization literature \cite{mahasseni2017unsupervised,yuan2019cycle, apostolidis2019stepwise,apostolidis2020ac}, where reconstruction fidelity was often leveraged implicitly. The underlying rationale holds that unique and salient frames are more critical for reconstructing the original video's content than redundant frames. Our prior work, RS-SUM \cite{myicip2023sum}, transitioned from an implicit principle to a direct measurement by training a generator to reconstruct videos from a summary. This approach empirically validated the strong correlation between frame-level reconstruction degradation and human-annotated scores. To infer importance, RS-SUM used iterative frame masking; however, this method lacked a dedicated summarizer and suffered from high inference costs. TR-SUM addresses RS-SUM's limitations with a dedicated single-pass scoring model and a new RL training pipeline. In TR-SUM, the generator is used exclusively during the reward generation stage, and its training is upgraded with a dynamic window masking strategy.}

{The principle linking reconstruction quality to information representativeness originates from prior literature~\cite{mahasseni2017unsupervised,yuan2019cycle, apostolidis2019stepwise}, where it was often implicitly leveraged based on the rationale that salient frames are more critical for reconstruction. Our prior work, RS-SUM \cite{myicip2023sum}, transitioned this to a direct measurement by training a generator to reconstruct videos from a summary and used iterative frame masking to empirically validate the correlation between reconstruction degradation and human scores. While effective, RS-SUM suffered from high inference costs. TR-SUM addresses this with a single-pass summarizer trained using (RL) and an upgraded generator for reward generation.}

These design changes form the foundation for TR-SUM’s core contributions, which are outlined below:}
\begin{itemize}
    \item {A generator-guided reward model that aligns summary quality with the frame-level contribution to reconstruction fidelity, providing a semantically grounded training objective.}
    \item {A two-stage training paradigm that separates generator and summarizer optimization, offering improved training stability over joint or adversarial setups.}
    \item {A dynamic masking strategy applied during generator pretraining. Instead of masking frames randomly, we use a rule-based scheme that preserves contextual continuity and improves reconstruction quality. This results in a stronger generator and improves both the initialization and downstream performance of the summarizer.}
    \item {A new initialization scheme and per-video baseline tracking for reinforcement learning.}
\end{itemize}

The rest of the paper is organized as follows. Section~\ref{sec:rel} reviews the related work on unsupervised video summarization. Section~\ref{sec:apr} describes the proposed method in detail. Section~\ref{sec:exp} presents the experimental results and analysis. {Section~\ref{sec:disc} discusses the design choices, and future directions.} Section~\ref{sec:con} concludes the presented work.

\section{Related Works}\label{sec:rel}

{Recent research has increasingly concentrated on multimodal unsupervised video summarization, incorporating diverse data types such as audio, video, captions, and user queries, often enhanced by contrastive learning techniques~\cite{wang2024m3sum,li2023progressive,huang2024aesthetic,narasimhan2021clip}. Some approaches even employ large language models (LLMs) to generate summaries~\cite{sugihara2024language,wang2024m3sum}. While these multimodal methods can improve summarization quality, models that focus exclusively on video content are generally more computationally efficient. For instance, the CLIP model used in~\cite{narasimhan2021clip, xu2023self} has 150 million parameters, compared to just 5 million parameters in~\cite{SummarizingACheat}. Additionally, multimodal methods may become inefficient when only the video frames are available and other modalities are missing. As an example, in~\cite{narasimhan2021clip}, the performance of the CLIP model is comparable to the frame-only method from~\cite{SummarizingACheat} when captions are unavailable, despite CLIP being 30 times larger. Given that our approach employs reinforcement learning with a reward-based mechanism, the following sections will explore unsupervised video summarization methods that emphasize video content and rely primarily on reinforcement learning for training.
}

Many unsupervised video summarization algorithms focused on video content follow the principle that a summary should allow viewers to understand the original content with less effort, time, and resources~\cite{mahasseni2017unsupervised, apostolidis2019stepwise, yuan2019cycle, jung2019discriminative, apostolidis2020ac, apostolidis2020unsupervised, jung2020global, liu2019learning, he2019unsupervised}. These algorithms use Generative Adversarial Networks (GANs) to generate a summary that captures the essence of the video. GAN-based unsupervised video summarization typically involves three units: a summarizer, a generator, and a discriminator. The summarizer assigns importance scores to frames, selecting high-scoring frames to form a summary. The generator creates two new video representations from the summary and the original video, aiming for similarity in content and style. The discriminator evaluates the generator’s outputs to determine which is based on the summary.

The adversarial learning process for training a keyframe selector using Long Short-Term Memory (LSTM) was first introduced by~\cite{mahasseni2017unsupervised}. Later works have focused on enhancing this approach with various modifications, such as developing a more robust discriminator to retain more information in a video’s summary~\cite{yuan2019cycle}, adjusting loss functions and optimization steps~\cite{apostolidis2019stepwise}, and adding a video decomposition stage, which breaks each video into smaller, non-overlapping chunks of consecutive frames with uniformly sampled strides before feeding them to the summarizer~\cite{jung2019discriminative}. Further refinements included incorporating a frame score refinement stage into the summarizer’s output~\cite{apostolidis2020unsupervised, apostolidis2020ac}, which involved an attention module that progressively edits each frame’s score based on current and previous frames~\cite{apostolidis2020unsupervised}, as well as the introduction of an Actor-Critic model that adjusts the frame scores in a non-sequential order, considering past and future frames and previous adjustments~\cite{apostolidis2020ac}.

{GANs have been applied to unsupervised video summarization due to their ability to generate diverse and realistic summaries. However, balancing and coordinating the training of these models to ensure convergence and stability is challenging~\cite{li2023unsupervised}. Issues such as mode collapse, vanishing gradients, and oscillations can lead to severe training instabilities~\cite{zhou2018deep, arjovsky2017towards}. Stabilizing GAN training for video summarization remains an active area of research. {SUM-SR addresses this by removing the discriminator~\cite{li2024unsupervised}, while recent works explore diffusion models as more stable alternatives~\cite{yu2024unsupervised}.} Other methods explored the integration of reinforcement learning with custom reward functions to have a more stable training process~\cite{zhou2018deep, gonuguntla2019enhanced, zhao2019property, yoon2021interp, yaliniz2021using,yuan2022unsupervised}.} A custom reward function measures specific properties required in an optimal video summary, such as diversity, representativeness, smoothness, and sparsity. A two-part reward function called Diversity-Representativeness was suggested in \cite{zhou2018deep} that measured diversity by examining differences between frames of the summarized video and representativeness by comparing the selected frames to the entire video. The aim was to train a model that created a summary of diverse and representative frames from different parts of the video. {Some methods emphasize shot-level semantics over frame-wise relations to improve performance~\cite{yuan2022unsupervised}, while others introduce multimodal rewards based on transcript content~\cite{barbakos2025unsupervised}. Although noteworthy, the latter depends on accurate transcripts and cannot be applied to standard benchmarks. A zero-shot approach~\cite{pang2023contrastive} replaces reinforcement learning with contrastive losses that capture frame uniqueness and global consistency. It selects informative frames effectively, but the lack of temporal modeling limits story coherence and results in low F-scores. CA-SUM~\cite{SummarizingACheat} removes reinforcement learning by directly optimizing attention for diversity and uniqueness. SegSum~\cite{vo2025integrate} extends this with temporal segmentation, but its inference is slow despite the model being lightweight.}

Many video summarization methods mentioned above use LSTM-based models and therefore could have problems such as vanishing and exploding gradients~\cite{li2018independently}. To address these challenges, some unsupervised video summarization methods have incorporated self-attention modules or transformer encoders~\cite{vaswani2017attention} into their LSTM-based models~\cite{jung2020global,liu2019learning,he2019unsupervised,vaswani2017attention}. Others have exclusively utilized transformer encoders~\cite{yoon2021interp, SummarizingACheat,vo2025integrate}. These strategies primarily concentrate on substituting LSTM-based models with self-attention encoders. Despite these modifications, these methods continue to employ reward-based training, utilizing traditional rewards such as representativeness/diversity and length regularization cost~\cite{zhou2018deep,apostolidis2019stepwise}.

{The training method in~\cite{zhou2018deep} provides an alternative to adversarial learning and has been widely adopted (e.g.,~\cite{zhou2018deep, gonuguntla2019enhanced, zhao2019property, yoon2021interp, SummarizingACheat, yaliniz2021using,vo2025integrate}). While the core framework of reward-based training remains unchanged in both prior works and ours, we introduce a fundamental innovation by completely re-imagining the reward generation step. Instead of relying on handcrafted heuristics, our method employs a learned reward function, offering a more effective approach. This shift improves performance and presents a new direction for reinforcement learning-based summarization, potentially reducing reliance on handcrafted reward functions in future work. Further details on our methodology are presented in the next section.}

\section{Approach}\label{sec:apr}
\begin{figure*}
   \begin{center}
  \includegraphics[width=0.73\textwidth]{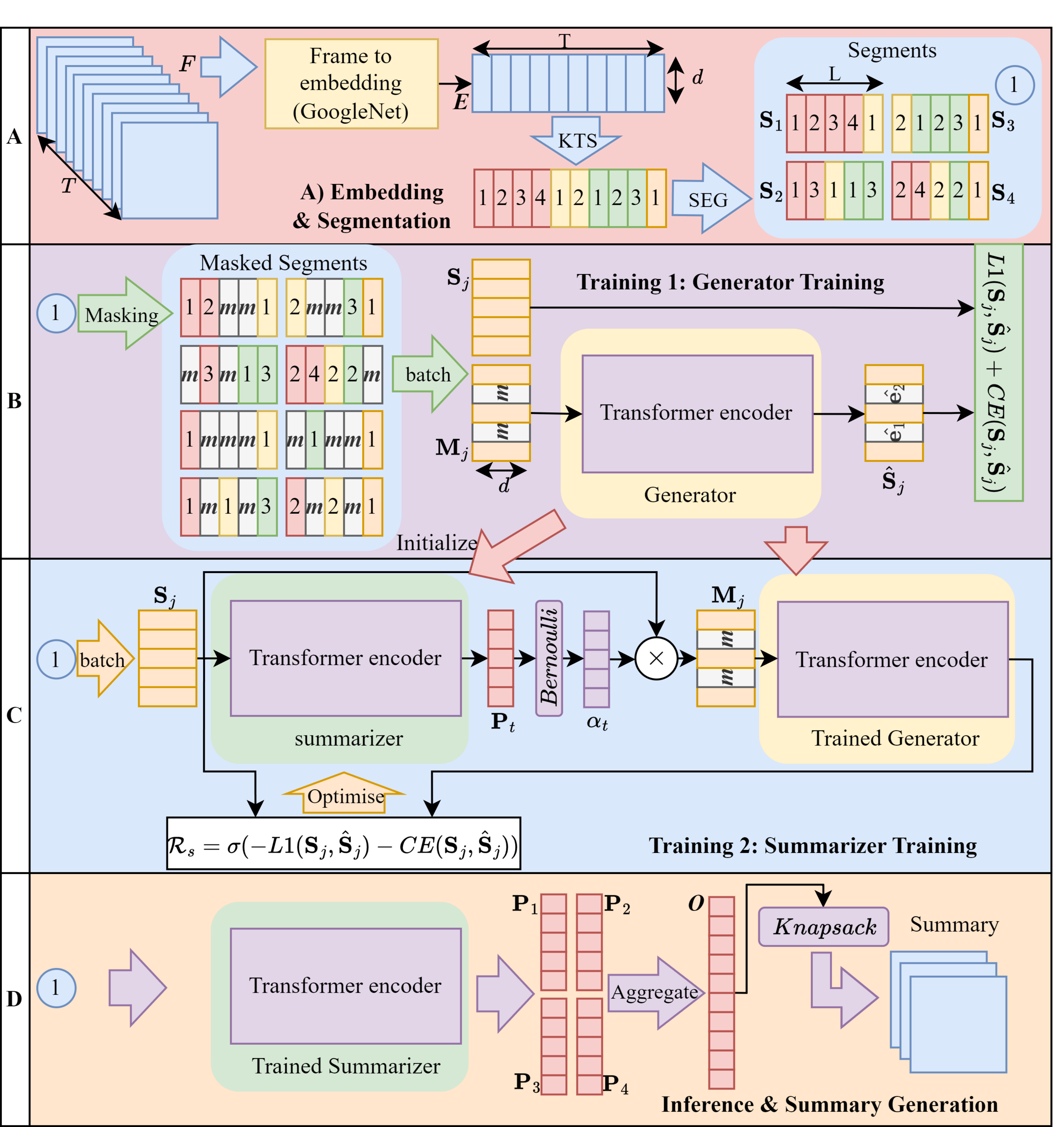} \caption{System flowchart: A) Input video is embedded, frames are shot labeled, and the sequence is broken down into segments. B) Segments are randomly masked based on the shot labels for self-supervised generator training. C) The trained generator is used to train the summarizer via reinforcement learning. D) The summarizer assigns scores to the embedding sub-sequences, which are combined to create a frame score sequence for the generated video summary.}\label{fig:all}
  \end{center}

\end{figure*}

Here, we introduce a novel method for generating and assessing video summaries using reinforcement learning that includes a learned reward function. Our method involves two models: a video generator and a video summarizer.

The input to the generator model is a masked video, in which some of the frames are masked (missing). The generator model takes this video summary as input and generates a reconstructed video. It learns to fill in for the missing frames by minimizing a reconstruction loss function that measures the similarity between the original and reconstructed frames.
% {The video generator is trained to reconstruct the original video from a provided summary by filling in any missing frames. The video summarizer is trained to identify the most crucial frames from a given video by maximizing the reward from the video generator. This reward is based on a reconstruction loss, which is a measure of the similarity between the original and reconstructed frames.}

% {The video generator model accepts a video summary as input and produces a reconstructed video. The generator model learns to fill in the missing frames by minimizing the reconstruction loss. The reconstruction loss is a measure of the similarity between the original and reconstructed frames.}
The video summarizer model takes a video as input and generates importance scores for each frame. It then creates a video summary using frames' importance scores.
% The video summarizer model takes a video as input and generates a series of importance scores, one for each frame.
To train the video summarizer, a video summary is first created by selecting the frames with the highest scores. Next, the summary video is passed to the generator for reconstruction. Finally, the reconstruction loss between the input video and its reconstruction is used to update the video summarizer. The video summarizer model learns to assign higher scores to frames that better represent the input video and contribute to a lower reconstruction loss. It is trained using reinforcement learning~\cite{sutton2018reinforcement}, where the reward is the sigmoid of the negative reconstruction loss.

{GAN-based methods often struggle with instability due to their adversarial training process, leading to challenges like mode collapse and training oscillations~\cite{li2023unsupervised,zhou2018deep,arjovsky2017towards}. To mitigate instability, we replace adversarial training with a decoupled training strategy for the generator and summarizer. This strategy allows each component to be optimized independently, leading to a more stable and reliable learning process. By preserving the key advantage of evaluating summaries through reconstructability, our approach retains the benefits of GAN-based methods without inheriting their drawbacks.}
% Unlike adversarial approaches, which demand careful balancing between competing objectives, our method provides a more structured and predictable optimization process. This not only improves training robustness but also eliminates reliance on handcrafted rewards.}

{The video summarizer model is built within an encoder-decomposition-summarizer-aggregation framework. In the following subsections, we describe each component of this framework and the training procedure. 
}
\begin{itemize}
\item Section~\ref{sec:dec} describes the encoding and decomposition steps, wherein the frames of the input video are converted into embeddings and the video is broken down into a smaller set of segments.
\item Section~\ref{sec:gen} outlines the architecture of the video generator and describes its training method.
\item Section~\ref{sec:sum} details the architecture of the summarizer and its training process.
\item Section~\ref{sec:inf} delineates the steps of the inference stage and the video summary generation pipeline.

\end{itemize}

In the rest of this paper, bold uppercase symbols like $\bm{I}$ stand for sequences, while bold lowercase elements such as $\boldE$ indicate vectors. Italic lowercase or uppercase letters represent numbers. For a glossary of all variable symbols, their types, and definitions, please refer to Table~\ref{tb:glos}.
\begin{table}[t] 
\setlength{\tabcolsep}{1pt}
  \centering
  \caption{{Glossary of variable symbols used in this paper}.}
\begin{tabular}{l|l}
\toprule
\multicolumn{1}{c|}{\textbf{Symbol}} & \textbf{Description} \\
\midrule
\multicolumn{2}{|c|}{Sequences} \\
\midrule
$\bm{F} \in \mathbb{R}^{T \times H \times W \times C }$ & Input video (frame sequence) \\
$\bm{E} \in \mathbb{R}^{T \times d}$ & Frame embeddings sequence \\
$\bm{S_{j/k}}\in \mathbb{R}^{L \times d}$ & $j$-th or $k$-th frame embedding sub-sequence \\
$\bm{M}_j\in \mathbb{R}^{L \times d}$ & Partially masked or summary sub-sequence \\
$\bm{\hat{S}}_j\in \mathbb{R}^{L \times d}$ & Reconstructed $j$-th frame embedding sub-sequence \\
$\bm{P}_j\in \mathbb{R}^T$ & $j$-th frame score sub-sequence (sequence of numbers) \\
$\bm{O}\in \mathbb{R}^T$ & Final frame score sequence (sequence of numbers) \\
$\bm{U}\in \{0, 1\}^T$ & User summary (binary sequence) \\
$\bm{A}\in \{0, 1\}^T$ & Automated summary (binary sequence) \\
\midrule
\multicolumn{2}{|c|}{Vectors} \\
\midrule
$\hat{\boldE}_{t/j} \in \mathbb{R}^{d}$& Reconstructed embedding of $t$-th or $j$-th frame \\
$\boldE_{t/j} \in \mathbb{R}^{d}$ & Embedding of $t$-th or $j$-th frame \\
$\boldM\in \mathbb{R}^{d}$ & Masked token embedding \\
$\boldH_{t}\in \mathbb{R}^{d}$ & $t$-th hidden state embedding \\
$\boldW_{sc}\in \mathbb{R}^{d}$ & Trainable weights of the scoring layer \\
\midrule
\multicolumn{2}{|c|}{Numbers} \\
\midrule
$T$ & A video's total number of frames \\
$L$ & Length of sub-sequences. \\
$J$ & Total number of sub-sequences \\
$\triangle$ & Sub-sequence starting point random shift \\
$d$ & Frame embedding dimension size \\
% $H$ & Compressed dimension size \\
$D_R$ & Dynamic window size ratio \\
$M_R$ & Total masking ratio \\
$p_{t}$ & $t$-th frame importance score \\
$a_{t}$ & $t$-th frame selection action \\
$N$ & Total number of episodes \\
$\mathcal{L}_{\text{CE}}$ & Cosine similarity loss \\
$\mathcal{L}_{\text{L1}}$ & L1 loss \\
$\mathcal{L}_{rec}$ & Reconstruction loss \\
$\mathcal{R}_s$ & Reward value \\
$b$ & Moving average of past rewards \\
$\mathcal{L}_{\text{reg}}$ & Regularization loss \\
$\delta$ & Regularization factor \\
$\beta$ & Regularization coefficient \\
$l$ & Number of transformer encoder layers \\
$h$ & Number of self-attention heads \\
$P$ & Precision \\
$R$ & Recall \\
$F$ & F-score \\
$H$ & frame height \\
$W$ & frame width \\
$C$ & frame channels \\
\end{tabular}%
% \caption*{\textit{Note:} $T$ = number of frames; $H$ and $W$ = frame height and width; $C$ = number of channels (e.g., 3 for RGB).}
  \label{tb:glos}%
\end{table}%

\subsection{Encoding and video segmentation}\label{sec:dec}
Fig.~\ref{fig:all}.A illustrates the workings of the encoder and decomposition stage. Consider a video comprising $T$ frames denoted as $\bm{F}$. The encoder, implemented as a Convolutional Neural Network (CNN), transforms the input video (frame sequence) $\bm{F}$ into the frame embeddings sequence $\bm{E}=\left\{\boldE_{t}\right\}_{t=1}^{T}$, where each $\boldE_{t} \in \mathbb{R}^{d}$ is the embedding representation of the $t$-th frame.
% {So you have T frames and the embedding length of each frame is also T?}. 
We employ GoogleNet~\cite{szegedy2015going} as the CNN model where the frame embeddings are the output of its penultimate layer. We opt for GoogleNet~\cite{szegedy2015going} to maintain consistency with most prior works~\cite{mahasseni2017unsupervised, apostolidis2019stepwise, yuan2019cycle, jung2019discriminative, apostolidis2020ac, apostolidis2020unsupervised,jung2020global,liu2019learning,he2019unsupervised, SummarizingACheat,li2024unsupervised,vo2025integrate} and to emphasize the impact of our algorithm on the results rather than the choice of the encoder. 
However, any arbitrary CNN such as~\cite{carreira2017quo} can be utilized in our proposed framework without loss of generality, as we do not make any specific assumption regarding the chosen feature encoder.

% However, other visual feature extractor models such as I3D~\cite{carreira2017quo} can also be utilized, which was the case in~\cite{gao2020unsupervised}.
% {which ones? you haven't introduced them yet} 

After obtaining the frame embeddings, we utilize Kernel Temporal Segmentation (KTS)~\cite{potapov2014category}, a method that divides a video into segments with minimized internal variance, to extract shot boundaries within a video.
% {within video or video segments?}.
Shots, in the context of video representation, represent continuous sequences of similar frames. In Fig.~\ref{fig:all}.A, some exemplar shots are shown, each with a different color, {where within each shot, frames are sequentially numbered starting from 1 up to the end of the shot.
As will be discussed in Section~\ref{sec:gen}, the obtained shots will be used in the next stage of the proposed approach.

% The outcome of the shot segmentation phase it is not required in this section. Instead, its usage details will be discussed in Section~\ref{sec:gen}.}
% {this sentence must be re-written: and the use of a number assigned to each frame within each shot,} which starts from 1 when the shot begins and counts up until the end of the shot. The concept of shots is utilized in the next step.

% {Please explain it simpler: why you use shots, why you use segments, and what is the relation between them?}

After obtaining the shot boundaries, each $\bm{E}$ is decomposed into a set of smaller video segments $\bm{S_j}=\{\boldE_{t}\}_{t=1}^{L}$, where $L$ is the video sub-sequence length and $j=1 \ldots J$ is the sub-sequence identifier. $J$ is the total number of segments, which is dependent on the input video's length {$T$}. Each video is divided into smaller segments using two methods: sequential split and dilated split. For sequential split, we select every $L$ consecutive frame as one sub-sequence. During training, we randomly shift the starting point of each sub-sequence by $\triangle$ (a value within the range of $\pm [0, L/2]$) for each training epoch to enhance the diversity of the samples. During inference, we do not apply any shift. For dilated split, we sample segments of $L$ frames with a variable dilation rate that depends on the input video length.
We start by padding the vector $\bm{E}$ with zeros until its length is divisible by $L$. If $n=\lceil T/L \rceil$, we then pick every $n$-th frame for each sub-sequence, starting from the first frame for the first sub-sequence, the second frame for the second sub-sequence, and so on.
% {Explain this further. How it works?}

% {What happened to the decoder? You didn't explain it.}

\subsection{Generator architecture and training}\label{sec:gen}
This section describes details of the proposed generator training stage. We refer to this stage as the self-supervised pre-training stage.
\begin{figure}
   \begin{center}
  \includegraphics[width=0.40\textwidth]{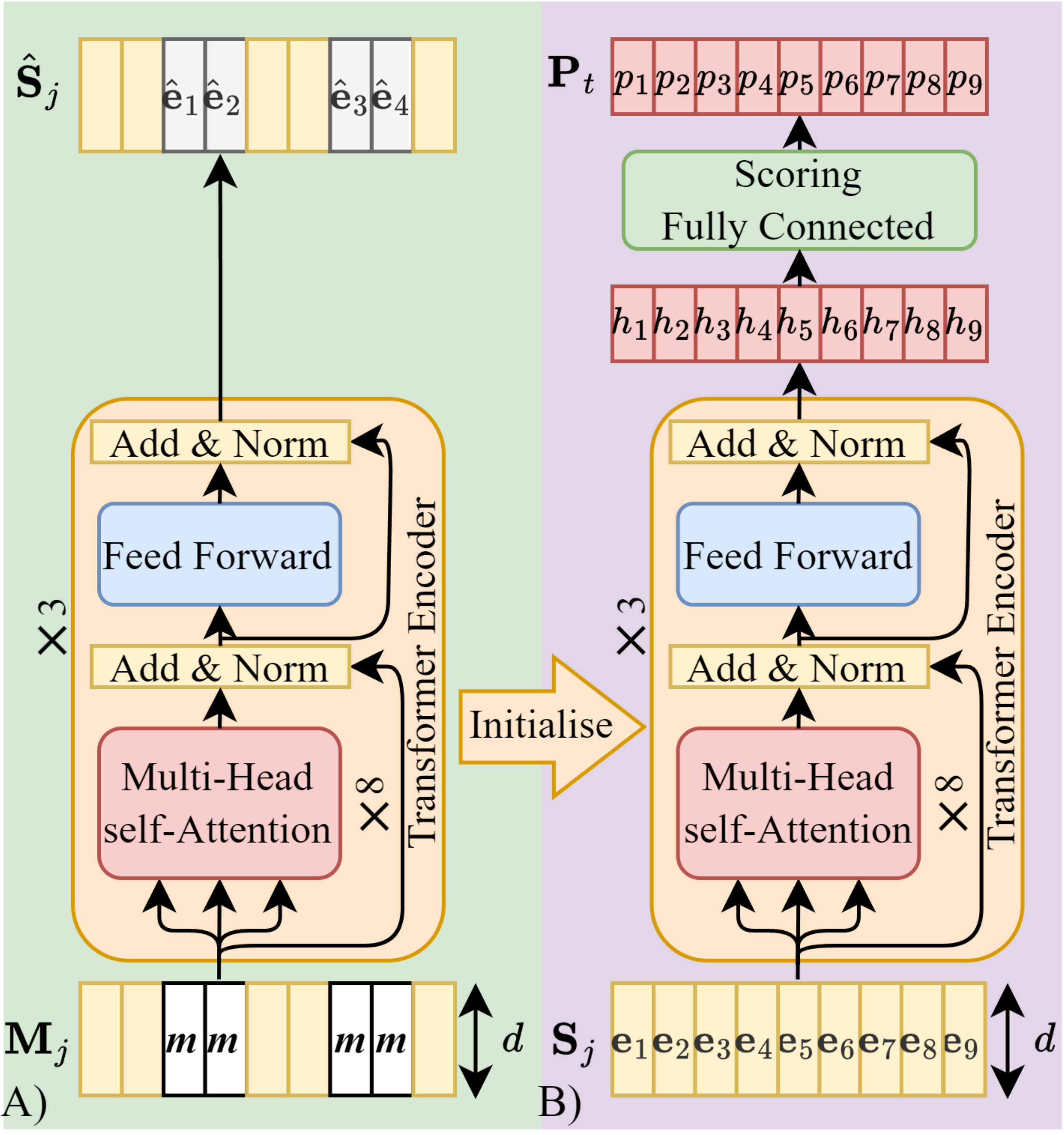} \caption{The architectures of A) generator B) summarizer. }\label{fig:arcsgensum}
  \end{center}

\end{figure}
{The generator is a transformer encoder, which comprises a stack of 3 identical layers~\cite{vaswani2017attention}}. The input to the video generator is a masked video sub-sequence, $\bm{M}_j$. The embedding of some frames in $\bm{M}_j$ are masked, meaning they are replaced with a special fixed masked token embedding {$\boldM$}, which is filled with arbitrary values.
The generator then tries to reconstruct the original embeddings at the masked frames using the embeddings of the non-masked frames to obtain a reconstructed video sub-sequence $\bm{\hat{S}}_j$. Fig.~\ref{fig:arcsgensum}.A shows the generator's architecture.

% preceded by an optional compression layer that can be used to reduce the model's computational complexity. The compression layer is a fully-connected layer with ReLU that reduces the dimensionality of each frame embedding, $\boldE_t$, from $d$ (e.g., 1024) to a smaller dimension, $H$(e.g., 512, 256, or 128). Following the transformer encoder is an expansion layer, which is another fully-connected (FC) layer with ReLU that increases the dimensionality of the frame embeddings from $H$ back to $d$ if the compression layer is used.

% {Please show the exact architecture of your video generator in a figure as it is one the main components of your work}

The generator training stage is depicted in Fig.~\ref{fig:all}.B. This generator does not require ground-truth annotations and uses the input video as the ground truth. It is trained in a self-supervised manner using the following loss function:
\begin{equation}\label{eq:ls}
\begin{split}
\mathcal{L}_{\text{CE}} &=\sum_{t=1}^{L}( 1 - \frac{{\boldE_t \cdot \hat{\boldE}_t}}{{\|\boldE_t\|_2 \cdot \|\hat{\boldE}_t\|_2}}), \\
\mathcal{L}_{\text{L1}} &= \frac{1}{L}\sum_{t=1}^{L} ||\boldE_t-\hat{\boldE}_t||_1,\\
\mathcal{L}_{rec}&=\mathcal{L}_{\text{CE}} + \mathcal{L}_{\text{L1}},
\end{split}
\end{equation}
where $\boldE_t$ stands for the $t$-th frame embedding of the input video sub-sequence, while $\hat{\boldE}_t$ denotes the reconstructed frame embedding. $\mathcal{L}_{rec}$ measures the similarity between the input video and its reconstructed version. The above loss function is a combined function that calculates both the absolute difference ($\mathcal{L}_1$) and the cosine similarity ($\mathcal{L}_{CE}$) between the reconstructed and the original frame embeddings (CE stands for cosine embedding loss).

% Specifically, the generator is trained to reconstruct the masked frames. For each input video sub-sequence $\bm{S}_j$, we mask a portion (several frames) of the sub-sequence by replacing them with a mask token embedding.

The masking algorithm {(illustrated in Algorithm \ref{alg:mask})} consists of two steps. In  the first step, a random shot of the video is selected. Within that shot, a random window of consecutive frames, with a length equal to $D_R$ (\%) of the shot length, is chosen as a window candidate. This process continues until the total sum of the window candidates' lengths reaches $M_R$ (\%) of $L$. In the second step, we apply masking to the window candidates. Each window candidate undergoes a flexible masking operation, which can result in one of the following three outcomes:
\begin{itemize}
    \item Masking: where each frame embedding within the window candidate is replaced by a masked token embedding $\boldM$. This has an 80\% chance of occurring in our setup.
    \item Replacement: where the window candidate is replaced with another randomly selected window that is not among the window candidates. This has a 10\% chance of occurring in our setup.
    \item No change: where the window candidate remains as is. This has a 10\% chance of occurring.
\end{itemize}
% The first outcome is masking, where each embedding of the window candidate is replaced by $\boldM$, the masking embedding. This has an 80\% chance of occurring. The second outcome is replacement, where the window candidate is replaced with another randomly selected window that is not already among the window candidates. This outcome has a 10\% chance of happening. The third and final outcome is no change, where the window candidate remains as it is. This also has a 10\% chance of occurring.
% Each window candidate has an 80\% chance of being masked, a 10\% chance of being replaced with another part that is not chosen {it is not clear that much} and has the same size (we call this "replacement"), and a 10\% chance of staying the same (we call this "No change"). This way, we have a dynamic masking ratio. 

\begin{algorithm}
\SetAlgoLined
\caption{{Masking Algorithm}}\label{alg:mask}
\textbf{Input:} Subsequence $\bm{S_j}=\{\boldE_{t}\}_{t=1}^{L}$, where $\bm{S_j}$ consists of multiple shots \\
\textbf{Definition:} A shot is a group of consecutive, similar frame embeddings $\{\boldE_{ss}, \dots, \boldE_{se}\}$ within $\bm{S_j}$, with $|se - ss|$ varying by shot \\
\textbf{Parameters:} Diversity ratio $D_R$, masking ratio $M_R$ \\
\textbf{Initialize:} Candidate Collection set (CC) $\leftarrow \emptyset$\\

\While{ $|\text{CC}|$ $< M_R \cdot L$}{
    Randomly select a not selected shot from $\bm{S_j}$\;
    Randomly select\\
    $\{\boldE_i, \dots, \boldE_{i+m}\}$ $\in$ shot $|$ $ m = \lfloor D_R \cdot \text{shot Length} \rfloor$\;
    $Add$ $\{\boldE_i, \dots, \boldE_{i+m}\}$ to CC\;
}
\For{$\{\boldE_i, \dots, \boldE_{i+m}\} \in \text{CC}$}{
    With probability 0.8, change each one to $\boldM$\;
    With probability 0.1, replace with \\
    randomly selected $\{\boldE_k, \dots, \boldE_{k+n}\} \notin$ CC\;
    With probability 0.1, make no change\;
}
\textbf{Output:} $\bm{M}_j$\\
\end{algorithm}

This window masking scheme, which distinguishes our approach from prior work~\cite{myicip2023sum}, serves two main purposes. First, by masking frames that are similar to the masked frame within the same shot, we facilitate the generator model to find more complex frame relations during the training process. 
Second, incorporating a dynamic ratio alongside partial shot masking ensures the model retains contextual information about the frames, preventing entirely blind reconstructions.

% we train the generator model to be better at finding complex frame relations and doing context matching instead of learning to interpolate frames. Second, by using a dynamic ratio and not masking the whole shot, we ensure that the model has some information about the context of the masked frames and that it does not make a completely blind reconstruction.

 % At this stage, the weights of the generator are used as the initial weights for the summarizer or decoder, which is detailed in the next section.

\subsection{Summarizer's architecture and training}\label{sec:sum}
The summarizer model {is a transformer encoder with a stack of 3 identical layers, followed by} a fully-connected (FC) layer with a sigmoid activation function. Fig.~\ref{fig:arcsgensum}.B illustrates the summarizer's architecture. Here, the same encoder module as in the generator is utilized where its weights are {initialized using the weights of the generator trained in Section~\ref{sec:gen}}. Fig.~\ref{fig:arcsgensum}.B illustrates the summarizer's architecture. {The key distinction between the generator and the summarizer is the added FC layer that maps each $d$-dimensional frame embedding into a single frame score. The FC layer is initialized randomly.}

The {summarizer}'s transformer encoder accepts the frame embeddings {sub-sequence}  $\bm{S_j}=\left\{\boldE_{t}\right\}_{t=1}^{L}$ as input and yields the hidden states $\left\{\boldH_{t}\right\}_{t=1}^{L}$ for each frame. These hidden states encapsulate the temporal dependencies and contextual information of the frames. The final FC layer then calculates a frame importance score or selection probability, $p_{t}$, for each frame, signifying its relevance to the generated summary as follows:
\begin{equation}\label{eq:pt}
p_{t}=\sigma\left(\boldH_{t}. \boldW_{sc}\right),
\end{equation}
{where $\boldW_{sc} \in \mathbb{R}^{d}$ represents the trainable weights of the FC layer. Essentially,  $\boldW_{sc}$ acts as a trainable parameter, related to a self-gating  mechanism~\cite{ramachandran2017searching} that determines the importance of each frame.}
% 
% This is illustrated in Eq.~\eqref{eq:pt}, where $W_{sc} \in \mathbb{R}^{H_{D} \times 1}$ denotes the weights of the fully-connected (FC) layer, and $\sigma$ represents the Sigmoid activation function.

% {You can also say that W is a trainable parameter which allows us to have a self-gating mechanism}

Fig.~\ref{fig:all}.C illustrates the summarizer training process. The training process is as follows: the summarizer takes in the input video sub-sequence $\bm{S}_j$ and generates $\left\{p_{t}\right\}_{t=1}^{T}$. A frame action sub-sequence $\left\{a_{t}\right\}_{t=1}^{L}$ is then generated by sampling each individual $p_t$, as follows:
\begin{equation}\label{eq:ber}
    a_{t} \sim \operatorname{Bernoulli}\left(p_{t}\right),
\end{equation}
where $a_{t} \in\{0,1\}$ indicates whether the $t$-th frame is selected or replaced with the masked token embedding ($\boldM$). The summary $\bm{M}_j$ is defined as:
\begin{equation}\label{eq:Malph}
\bm{M}_j = \{\boldE_t \text{ if } a_t = 1 \text{ else } \boldM, \, t = 1, 2, \ldots T\}.
\end{equation}
The summary $\bm{M}_j$ is then passed to the video generator, which reconstructs the input video. The reconstruction loss in~\eqref{eq:ls} between the reconstruction and the original is calculated and converted into a reward ($\mathcal{R}_s$) using the following equation:
\begin{equation}\label{eq:rs}
\begin{split}
\mathcal{R}_s&=\sigma(-\mathcal{L}_{rec} ).
\end{split}
\end{equation}
% where $\boldE_j$ stands for the $j$-th frame embedding of the input video sub-sequence, while $\hat{\boldE}_j$ denotes the reconstructed frame embedding.
Which indicates that $\mathcal{R}_s$ is equal to the sigmoid of the negative value of the reconstruction loss. {During training, the goal of the summarizer is to increase $\mathcal{R}_s$ over time, which based on~\eqref{eq:rs} happens when $\mathcal{L}_{rec}$ is minimized. In essence, the summarizer is trained to create summaries that enhance the quality of the video reconstruction, focusing on the similarity between the original and the reconstructed frame embeddings.} {This formulation is an engineering choice for training stability. The negated loss aligns the learning objective, while the sigmoid function acts as a 'soft clipping' mechanism. This bounds the reward signal to prevent extreme values from destabilizing the agent's policy updates, ensuring smoother convergence.}

Mathematically, the summarization's objective is to learn a policy function~\cite{sewak2019policy}, denoted as $\pi_{\theta}$, with parameters $\theta$. This function optimizes the anticipated rewards and is defined as follows:
\begin{equation}
J(\theta)=\mathbb{E}_{p_{\theta}\left(\left\{a_{t}\right\}_{t=1}^{L}\right)}[\mathcal{R}_s],
\end{equation}
where $J(\theta)$ is the objective function. The goal is to find parameter values that maximize the objective function. The probability of a sequence of actions $\left\{a_{t}\right\}_{t=1}^{L}$ under the policy parameterized by $\theta$ is represented by $p_{\theta}(\left\{a_{t}\right\}_{t=1}^{L})$. In this context, actions represent the choice of whether to keep a frame in the summary or not. The expectation operator is represented by $\mathbb{E}$. 
% The reward function, denoted as $\mathcal{R}_s$ and defined by eq.~\eqref{eq:rs}, offers a reward for a state-action pair. In this context, a state-action pair refers to the selection of a set of frames for the summary, given the input video.
% This equation forms the foundation of our reinforcement learning approach for video summarization. The aim is to learn a policy that selects the most important frames to include in the summary, such that the expected cumulative reward is maximized. The reward is a function of the quality of the summary, as measured by the reconstruction loss when the original video is reconstructed from the summary.

% In this context, $p_{\theta}\left(a_{1: L}\right)$ signifies the probability distributions over all possible action sequences. The reward $R_s$ is determined by Eq.~\eqref{eq:rs}. 
Using the episodic REINFORCE algorithm~\cite{sutton2018reinforcement}, we can calculate the derivative of the objective function $J(\theta)$ with respect to $\theta$~\cite{williams1992simple}: 
% \begin{equation}
% \nabla_{\theta} J(\theta) \approx \frac{1}{N} \sum_{n=1}^{N} \sum_{t=1}^{L} \mathcal{R}_s^{(n)} \nabla_{\theta} \log \pi_{\theta}\left(a_{t} \mid \boldH_{t}\right),
% \end{equation}
\begin{equation}\label{eq:n}
\nabla_{\theta} J(\theta) \approx \frac{1}{N} \sum_{n=1}^{N} \sum_{t=1}^{L}\left(\mathcal{R}_s^{(n)}-b\right) \nabla_{\theta} \log \pi_{\theta}\left(a_{t} \mid \boldH_{t}\right).
\end{equation}
where $\mathcal{R}_s^{(n)}$ represents the reward at the $n$-th episode, and $N$ denotes the total number of episodes. {Here, an episode refers to each iteration of sampling $p_t$, creating $\bm{M_j}$, and then calculating its reward.} {The baseline $b$ is a moving average element incorporated into the policy gradient to ensure that updates focus on relative improvements. By centering the advantage function around zero, this baseline effectively reduces gradient variance~\cite{greensmith2004variance}, which improves training stability~\cite{mao2018variance}. {It is initialized as the sigmoid function applied to the negative of the generator's average loss at the end of training}, which facilitates faster convergence. For each training video, a separate instance of the baseline is maintained and updated solely based on the new rewards from subsequences generated for that video, as defined  below:
}

\begin{equation}\label{eq:b}
{b = 0.9 b + 0.1 \mathcal{R}_s^{(n)}}.
\end{equation}
% To facilitate convergence and reduce variances, we subtract the moving average of past rewards, $b$, from the reward~\cite{sutton2018reinforcement}. The updated gradient is:

% Considering that choosing a greater number of frames generally results in an improved $R_s$, we incorporate a regularization term. This term below is designed to restrict the number of selected frames. This ensures that the model doesn't artificially inflate the reward by simply selecting an increasing number of frames. $\delta$ is called the regularization factor.
The summarizer can inflate $\mathcal{R}_s$ by assigning a higher score to frames with fewer masked frames, making them easier to reconstruct by the generator. To counteract this, we employ a regularization loss, $L_{\text{reg}}$, defined as:
\begin{equation}\label{eq:reg}
\mathcal{L}_{\text{reg}} = |\frac{1}{T} \sum_{t=1}^{T} p_{t} - \delta|,
\end{equation}
where $\delta$ is the regularization factor. This regularization loss restricts the number of frames chosen for the summary by aligning the average frame scores with $\delta$. This approach imposes sparsity in the frame selection. 
% This regularization term limits the number of frames selected for the summary by centering the average frame scores around $\delta$ and introduces sparsity and diversity in the frame selection. 
% {why the model may collapse?}. We opt for $L1$ regularization because it introduces sparsity and diversity in the frame selection. Conversely, $L2$ regularization would have resulted in frames all close to $\delta$, the regularization factor, thereby limiting the diversity and dynamic range of scores in the model output.

% {I know the idea of this regulatization term but the reader may not figure it out based on your text. Also why L1 is used?}

We employ the stochastic gradient method to train the proposed model. Specifically, the model parameters are updated as follows:
\begin{equation}\label{eq:sgd}
\theta = \theta - \gamma \nabla_{\theta}\left(-J + \beta \mathcal{L}_{\text{reg}}\right),
\end{equation} 
where the calculated gradient and the regularization term are combined. Here, $\gamma$ represents the learning rate, and $\beta$ is the coefficient for the regularization factor.

\subsection{Inference and summary generation}\label{sec:inf}
The inference stage is illustrated in Fig.~\ref{fig:all}.D. During this stage, the input frame embeddings $\bm{E}=\left\{\boldE_{t}\right\}_{t=1}^{T}$ are first divided into multiple segments $\bm{S}_k=\{\boldE_{t}\}_{t=1}^{L}$, where $L$ is the new video length and $k=1 \ldots K$ is the sub-sequence identifier. This process is carried out using the operations described in Section~\ref{sec:dec} with $\triangle$ set to 0. This parameter, $\triangle$, was initially developed for training to introduce diversity in the training data. However, during inference, such diversification is unnecessary, hence the decision to set it to 0. Each $\bm{S}_k$ is then passed to the summarizer, which produces a frame score sub-sequence, $\bm{P}_k=\left\{p_{t}\right\}_{t=1}^{L}$, where each $p_t$ indicates the importance score of each frame. However, since the input video was decomposed into multiple overlapping segments, each frame receives multiple importance scores. We compute the final frame score $o_{t}$ for a single frame by calculating the average of these assigned scores for a single frame. The final output of this frame score generation algorithm is $\bm{O}=\left\{o_{t}\right\}_{t=1}^{T}$, which is the sequence of all final frame scores. This pipeline is presented in Algorithm~\ref{alg:fsg}.

\begin{algorithm}
\SetAlgoLined
\KwIn{$\bm{E}$ = \(\{\boldE_1,\:\boldE_2,\: ...\boldE_T\}\)}
% Counter: $\boldC \leftarrow [c_1,\:c_2,\: ...c_T] \leftarrow [0,\:0,\:0,\:...]$\;
% Final Scores: $\boldO$\( \leftarrow [o_1,\:o_2,\: ...o_T] \leftarrow [0,\:0,\:0,\:...] \)\;
$\bm{E}$ is split into multiple segments \( \{\bm{S_1},\:\bm{S_2},\: ...\}\)\;
% $\mathbf{G}$ \( \leftarrow [\bm{S_1},\:\bm{S_2},\: ...]\)\; 
\For {$\bm{S}_k$ $\in$ $\{\bm{S_1},\:\bm{S_2},\: ...\}$}{
 $\bm{P}_k$ $\leftarrow$ Summarizer($\bm{S}_k$)\;
% \For {$p_{t \in \{1..T\}} \in$ $\boldP_k$}{
% $c_t \leftarrow c_t + 1$\;
% $o_t \leftarrow o_t + p_t$\;
% }
}
\For {$t \in \{1..T\}$}{
$o_t \leftarrow Average(p_{t} \in \bm{P}_{k \in \{1..K\}} )$\;
}
\KwOut{$\bm{O} \leftarrow \{o_1,\:o_2,\: ...o_T\}$}
 \caption{Frame Score Generation Algorithm}\label{alg:fsg}
\end{algorithm}

We set the summary length limit to 15\%, which is a typical and commonly used number~\cite{surv}. Most methods for generating summaries select the most informative shots from a video. The informativeness of a shot is calculated by averaging the scores of all its frames (shot-level score). The goal is to select as many high-scoring shots as possible without exceeding the summary length limit. This selection step can be considered as a binary Knapsack problem, which can be solved using dynamic programming~\cite{TVSUM}. The final video summary is the solution obtained from this process.
% Most summary generation methods select the most informative shots from a video to create summaries. A shot’s informativeness is computed by averaging the scores of all the frames in that shot (shot-level score). The task is to select as many high-scored shots as possible without exceeding the summary's length limit. Normally, shorter shots with high scores are more likely to be selected in this process. The usual rule for video summary generation is that the summary should be between 5\% and 15\% of the original video’s length. Therefore, we set the length limit for the summary to 15\%. For a summary generation, the goal is to select as many high-scored shots as possible without exceeding the summary length limit. This selection step is a 0/1 Knapsack problem optimally solved using dynamic programming~\cite{song2015tvsum}. The final video summary is the solution that we obtain from this problem.

% As previously mentioned, the overall structure of the model follows an encoder-decomposition-summarizer-aggregation framework. To generate frame scores for a video, the video is first passed through the encoder and decoder, as shown in Fig.~\ref{fig:all}.A, to be converted into frame embeddings and decomposed into segments of length $L$ with $\triangle$ set to 0. Subsequently, these segments are passed to the decoder/summarizer to generate a frame score for each sub-sequence. Finally, the scores of all segments are aggregated to create the output frame score sequence, which is equal in length to the input video.

\subsection{Relative Comparison with Prior Works}

{Compared to RS-SUM~\cite{myicip2023sum}, our method introduces a single-pass summarizer that eliminates the need for iterative scoring. We further improve generator training by replacing the masking strategy with a dynamic window-based scheme, which better preserves temporal context and improves reconstruction quality. To enhance training stability, we adopt a two-stage optimization framework that separates generator and summarizer training, in contrast to prior adversarial approaches~\cite{mahasseni2017unsupervised, apostolidis2019stepwise, yuan2019cycle, jung2019discriminative, apostolidis2020ac, apostolidis2020unsupervised, jung2020global, liu2019learning, he2019unsupervised}. Unlike previous RL-based methods~\cite{zhou2018deep, gonuguntla2019enhanced, zhao2019property, yoon2021interp, SummarizingACheat,yuan2022unsupervised}, our reward function is derived from reconstruction fidelity, providing a closer alignment with human judgment. We also introduce per-video baseline tracking, initialized from the generator’s final epoch loss.}

\section{Experimental Results}\label{sec:exp}
In this section, we present and discuss our experimental results, compare them with the current SOTA methods, and conduct an ablation study. To ensure a fair comparison, we followed a widely accepted evaluation protocol, and used the same datasets and evaluation methods utilized by many leading approaches~\cite{mahasseni2017unsupervised, apostolidis2019stepwise, yuan2019cycle, jung2019discriminative, apostolidis2020ac, apostolidis2020unsupervised,jung2020global,liu2019learning,he2019unsupervised, SummarizingACheat} in this field. Subsequent sections will provide detailed insights into this standardized procedure and a comprehensive presentation of our findings.
\subsection{Datasets and the evaluation method}
To evaluate the performance of our proposed method, we utilized two standard benchmark datasets: SumMe~\cite{SUMME} and TVSum~\cite{TVSUM}. {These datasets validate quality against collective human judgment by using a consensus from multiple annotators. Strong performance, therefore, inherently demonstrates alignment with this standard.} The TVSum dataset consisted of 50 videos ranging from 1 to 11 minutes in duration. These videos were annotated by 20 users, who assigned an importance score on the scale of 1 to 5 to each 2-second frame sub-sequence.
Conversely, the SumMe dataset consisted of 25 videos with durations spanning 1 to 6 minutes.
The annotation process was performed by 15 to 18 individuals who created a summary for each video by selecting key (the most important) shots within each video. These summaries had to be between 5\% and 15\% of the total video length.

The predominant evaluation metric employed in SOTA video summarization methods is the F-Score similarity measure~\cite{mahasseni2017unsupervised, apostolidis2019stepwise, yuan2019cycle, jung2019discriminative, apostolidis2020ac, apostolidis2020unsupervised,jung2020global,liu2019learning,he2019unsupervised, SummarizingACheat}. F-Score quantifies the similarity between the automatically generated video summary and the user-annotated summary by assessing the overlap between the user summary ($\bm{U}$) and the automated summary ($\bm{A}$), both of which are sequences of 0s and 1s representing not selected or selected frames of the summary. The formula for calculating the F-Score is as follows:
\begin{equation} \label{eq:f}
F= 2\times 100 \times \frac{P \times R}{P+R} \%, \
P=\frac{\bm{A}\:\cap\:\bm{U}}{Len\left(\bm{A}\right)}, \
R=\frac{\bm{A}\:\cap \:\bm{U}}{Len\left(\bm{U}\right)},
\end{equation}
where $P$ and $R$ denote Precision and Recall, while $Len(.)$ is a function that returns the length of its input sequence.
For each video in the SumMe and TVSum datasets, the output of the video summarization algorithm was compared against annotations provided by all users. This comparison yielded multiple F-score values corresponding to the number of annotations. To consolidate these into a singular F-score, a reduction operation was applied. For TVSum, the established benchmark criterion involves averaging all F-scores to derive the final result. In contrast, for SumMe, the ultimate F-score was determined by selecting the maximum F-score among all evaluations. After obtaining the F-scores for all videos in each dataset, an overall F-score was computed for each video summarization method by averaging the F-scores across all videos in that dataset.

In~\cite{otani2019rethinking}, an evaluation method was introduced that compares frame-level automated scores with user-annotated frame importance scores using Kendall’s $ \tau $~\cite{kendall1945treatment} and Spearman’s $ \rho$~\cite{zwillinger1999crc} rank correlation coefficients. This approach was exclusively applicable to the TVSum dataset only, as the annotations for this dataset include frame-level importance scores assigned by users, which were not available in the SumMe benchmark. Both $\tau$ and $\rho$ were measures of rank correlation. In this context, the frame scores assigned by users and generated by the machine acted as the rankings. These two measurements were used to assess the similarity between these rankings.
This method held an advantage over the F-score measurement method as it was not influenced by the video shot segmentation mechanism.

For a given test video, the estimated frame-level importance scores were compared against the available user annotations. $ \tau $ and $ \rho$ values for each comparing pair are then computed. These values are then averaged to form the final $ \tau $ and $ \rho$ values for that test video. The computed $ \tau $ and $ \rho$ values for all test videos are then averaged, and this average is used to measure the method's performance on the test set.

To maintain consistency, we employed predefined 5-fold data splits (80\% training, 20\% test) for each dataset proposed by~\cite{zhang2016video} and used in~\cite{mahasseni2017unsupervised, apostolidis2019stepwise, yuan2019cycle, jung2019discriminative, apostolidis2020ac, apostolidis2020unsupervised,jung2020global,liu2019learning,he2019unsupervised, SummarizingACheat}. The experiment was replicated five times, once for each split, and the average results were reported here.

\subsection{Implementation setup}
%$D_R$ and $M_R$ are set to 50\% and 25\%, respectively. \textbf{}
In line with the standard approach adopted by SOTA unsupervised video summarization methods, we employed a pre-existing video feature extraction setup proposed by~\cite{zhang2016video}. In this setup, the feature arrays were generated through a two-step process: first, the input videos were downsampled to 2 frames per second (fps), and second, the 1024-dimensional output of GoogleNet's~\cite{szegedy2015going} penultimate layer was obtained for the sampled frames. During our proposed segmentation phase, videos were segmented into segments of $L=128$ frames. 

The architectural configuration of the proposed model was set as follows: The number of transformer encoder layers ($l$), the number of attention heads ($h$), and the hidden input dimension size $h$ were set to 3, 8, and 1024, respectively. The feedforward layers of the encoder had an expansion factor of 4, resulting in a hidden-state-dimension size of 4096. The scoring layer was an FC layer with an input dimension size of 1024 ($d$) and an output dimension size of 1.

The initial training phase, known as self-supervised training, spanned 250 epochs with a batch size of 128. The optimization was carried out using the AdamW optimizer in conjunction with a Cosine learning rate scheduler. The scheduler included a warm-up period of 100 epochs, during which the learning rate linearly increased from 0 to 0.01. Subsequently, the learning rate followed a cosine wave pattern, gradually decreasing after the warm-up period, to reach zero by the 1000th epoch. However, in our experiments, training was completed at epoch 250, before the learning rate reached zero. The training parameters of this phase including $D_R$ (dynamic masking ratio) and $M_R$ (sub-sequence masking ratio) were set to $0.5$ and $0.25$, respectively.  

Moving on to the summarizer training stage, we conducted 100 epochs with a batch size of 16. The parameter $N$ (number of episodes) was set to 5, and the learning rate was fixed at $0.00001$, utilizing the AdamW optimizer. The checkpoint with the least reconstruction loss was retained as the final model. In this phase, the parameter $\delta$ in~\eqref{eq:reg} was set to $0.5$. Additionally, $\beta$, the regularization loss coefficient, was set to $0.001$.
% The training parameters during this phase included $\delta$ as the regularization factor, set to $0.5$, and $\beta$ as the regularization loss coefficient, set to $0.001$.

Our experiments were executed on a Compute Canada node equipped with an NVIDIA V100 Volta GPU, with 32G HBM2 memory.
\subsection{Comparison against the state-of-the-art methods}
\begin{table}[t] 
\begin{threeparttable}
\renewcommand{\arraystretch}{0.8}
\setlength{\tabcolsep}{4pt}

  \centering
  \caption{{F-score comparison results.}}
% \begin{tabular}{l|c|c|c|c|c|c}
% \toprule
% Dataset & \multicolumn{2}{c|}{SumMe} & \multicolumn{4}{c}{TVSum} \\
% \midrule
% Method & F-score & Rank & F-score & Rank & \multicolumn{1}{l|}{$\tau$} & \multicolumn{1}{l}{$\rho$} \\
% \midrule
% SUM-GAN~\cite{mahasseni2017unsupervised} & 41.7 & 9 & 56.3 & 10 & –  & –  \\
% Cycle-Sum*~\cite{yuan2019cycle} & 41.9 & 8 & 57.6 & 8 & –  & –  \\
% DR-DSN*~\cite{zhou2018deep} & 41.4 & 10 & 57.6 & 8 & 0.02 & 0.026 \\
% SUM-GAN-AAE*~\cite{apostolidis2019stepwise} & 48.9 & 6 & 58.3 & 7 & –  & –  \\
% SUM-GAN-sl*~\cite{apostolidis2020unsupervised} & 47.8 & 7 & 58.4 & 6 & –  & –  \\
% CSNet~\cite{jung2019discriminative} & 51.3 & 3 & 58.8 & 5 & 0.025 & 0.034 \\
% AC-SUM-GAN*~\cite{apostolidis2020ac} & 50.8 & 5 & 60.6 & 4 &   &  \\
% CA-SUM*~\cite{SummarizingACheat}  & 51.1 & 4 & 61.4 & 2 &   &  \\
% RS-SUM*~\cite{myicip2023sum} & 52.5 & 2 & 61.4 & 2 &  \textbf{0.080} &  \textbf{0.106}\\
% TR-SUM & \textbf{54.5} & 1 & \textbf{62.3} & 1 & 0.078 & 0.104 \\
% \midrule
% Human & 54 & –  & 54 & –  & 0.177 & 0.204 \\
% \bottomrule
% \end{tabular}%

\begin{tabular}{l|c|c|c|c}
\toprule
Dataset & SumMe & \multicolumn{3}{c}{TVSum} \\
\midrule
Method & Fscore (SD) & Fscore  (SD) & $\tau$ & $\rho$  \\
\midrule
DR-DSN*~\cite{zhou2018deep} & 41.4 (1.2) & 57.6 (0.7) & 0.020 & 0.026 \\
SUM-GAN~\cite{mahasseni2017unsupervised} & 41.7 & 56.3 & –  & –  \\
Cycle-Sum*~\cite{yuan2019cycle} & 41.9 & 57.6 & –  & –  \\
EDSN~\cite{gonuguntla2019enhanced} & 42.6 & 57.3 & –  & –  \\
PCDL~\cite{zhao2019property} & 42.7 & 58.4 & –  & –  \\
{SB2S3}~\cite{ma2020similarity} & 42.8 & 57.8 & –  & –  \\
{Pang et al.~\cite{pang2023contrastive}} & 47.2 & 58.4 & 0.133 & 0.177 \\
Interp-SUM~\cite{yoon2021interp} & 47.7 & 59.1 & –  & –  \\
SUM-GAN-sl*~\cite{apostolidis2019stepwise} & 47.8 & 58.4 & –  & –  \\
SUM-GAN-AAE*~\cite{apostolidis2020unsupervised} & 48.9 (1.4) & 58.3 (0.6) & –  & –  \\
CSNet~\cite{jung2019discriminative} & 51.3 & 58.8 & 0.025 & 0.034 \\
{SUM-SR}~\cite{li2024unsupervised} & 51.3 & 60.2 & –  & –  \\
SUM-Ind~\cite{yaliniz2021using} & 51.4 (1.5) & 61.5 (1.7) & –  & –  \\
DMFF~\cite{yu2024unsupervised} & 53.0 (2.7) & 61.2 (0.8) & –  & –  \\
SLS~\cite{yuan2022unsupervised} & 52.0  & 62.0 & –  & –  \\
{SegSum*}~\cite{vo2025integrate} & 54.0 (1.6) & 62.0 (0.3) & –  & –  \\
RS-SUM~\cite{myicip2023sum} & 52.0 \textbf{(0.2)} & 61.1 (0.6) & 0.080 & 0.106 \\
\textbf{TR-SUM} & \textbf{54.5} (0.6) & \textbf{62.3 (0.3)} & \textbf{0.092} & \textbf{0.122} \\
\midrule
AC-SUM-GAN**~\cite{apostolidis2020ac} & 50.8 (1.9) & 60.6 (1.0) & 0.038 & 0.050 \\
CA-SUM**~\cite{SummarizingACheat}  & 51.1 (1.6) & 61.4 (1.1) & 0.160 & 0.210 \\
\midrule
Human & 54.0 & 54.0 & 0.177 & 0.204 \\
\bottomrule
\end{tabular}%

  \label{tb:soc}%
\begin{tablenotes}
    \item * Different $\delta$ for each dataset, **  different $\delta$ for each dataset fold.
\end{tablenotes}
\end{threeparttable}
\end{table}%
In this section, we present a comparative analysis between the outcomes produced by our approach, referred to as Trained Reward Summarizer (TR-SUM), and the current SOTA methods in unsupervised video summarization. 
\subsubsection{Quantitative Performance Comparison}\label{sec:exp1}

{In Table~\ref{tb:soc}, methods marked with $*$ used different regularization factors ($\delta$) for each dataset, selecting the $\delta$ that yielded the highest average F-score per fold (e.g., SumMe peaks at 0.4, TVSum at 0.7). Methods marked with $**$ averaged F-scores across varying $\delta$ values within folds (e.g., SumMe used 0.3 in the first fold and 0.6 in the second). In contrast, our method consistently used $\delta = 0.5$ for all folds and datasets.

Methods that used different $\delta$ values for each dataset~\cite{yuan2019cycle,zhou2018deep,apostolidis2019stepwise,apostolidis2020unsupervised} or folds~\cite{apostolidis2020ac,SummarizingACheat} introduced bias into the reported results. This is because the value of $\delta$ directly impacts the F-score by influencing the balance between frame scores and shot lengths. The Knapsack algorithm selects shots based on the ratio of shot frame scores and shot length. Varying $\delta$ shifts this distribution, impacting the shot score-to-length ratio differently across folds. To illustrate, Fig.~\ref{fg:deltaR} shows an experiment where frame scores were randomly generated from Gaussian distributions with 
$\delta$ values from 0.5 to 0.9. Repeating this five times per split, an optimal $\delta$ per split achieved an F-score of 59.2, while a constant $\delta = 0.5$ yielded 58, a 1.2-point artificial increase.
% To demonstrate this effect, we performed an experiment shown in Fig.~\ref{fg:deltaR}, where frame scores were randomly generated from Gaussian distributions with $\delta$ values ranging from 0.5 to 0.9. The process was repeated five times per data split, and the average results were recorded. By selecting an optimal $\delta$ per split, an F-score of 59.2 was achieved, while using a constant $\delta$ of 0.5 yielded 58, a 1.2-point spurious increase.}
\begin{figure}[t]
   \begin{center}
  \includegraphics[width=0.35\textwidth]{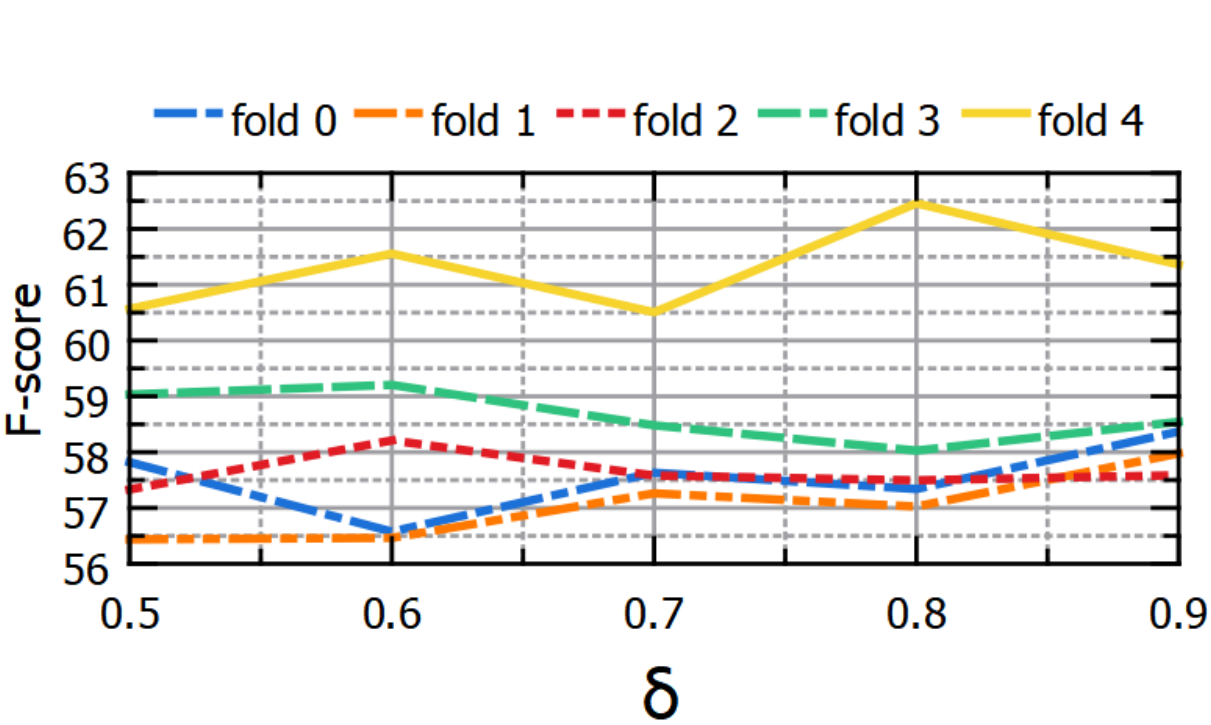} \caption{{Simulating the effect of different $\delta$ on each fold's F-score using Gaussian-sampled frame scores with varying median centers representing $\delta$.}}\label{fg:deltaR}
  \end{center}

\end{figure}

\begin{figure}[t]
   \begin{center}
  \includegraphics[width=0.49\textwidth]{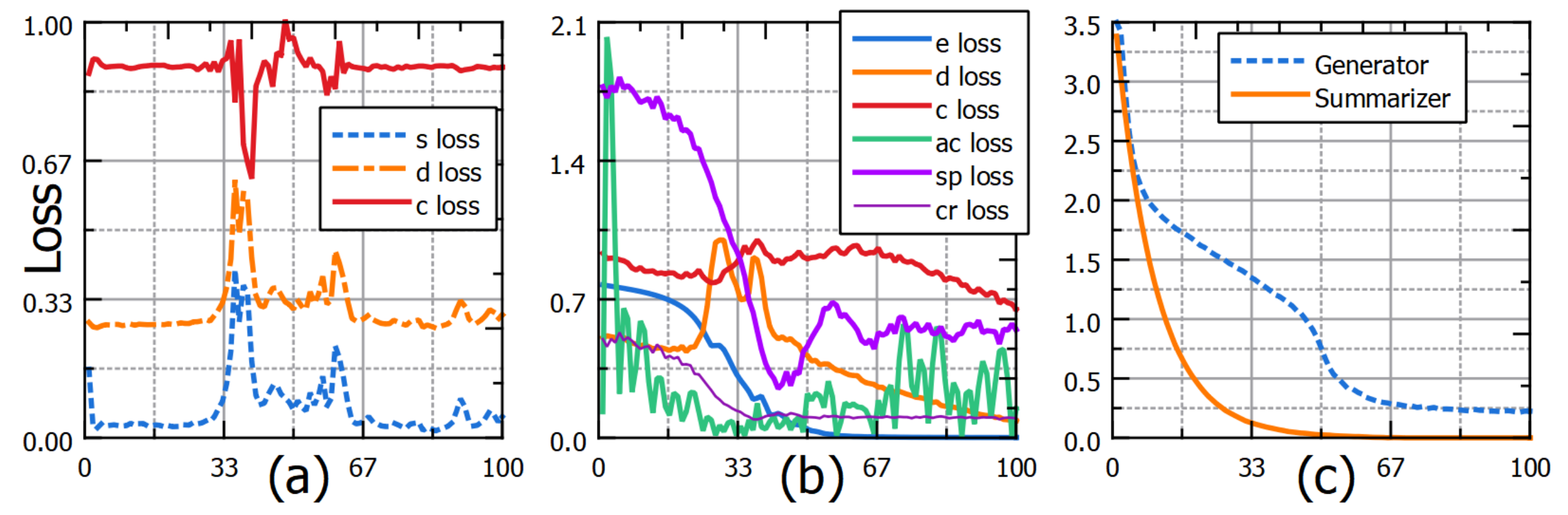} \caption{{Comparison of training loss curves for (a) SUM-GAN-AAE~\cite{apostolidis2020unsupervised}, (b) AC-SUM-GAN~\cite{apostolidis2020ac}, and (c) TR-SUM over 100 epochs on TVSum dataset.}}\label{fg:loss}
  \end{center}

\end{figure}
% \begin{figure}[t]
%    \begin{center}
%   \includegraphics[width=0.49\textwidth]{Figs/spacyFS.png} \caption{{Comparison of training loss and test F-scores (Fs) for (a) SUM-GAN-AAE~\cite{apostolidis2019stepwise}, (b) AC-SUM-GAN~\cite{apostolidis2020ac}, and (c) TR-SUM over 100 epochs.}}\label{fg:loss}
%   \end{center}

% \end{figure}

{The results in Table~\ref{tb:soc} show that our method achieves the highest F-score on both datasets, marking a considerable improvement over non-deep learning approaches such as~\cite{ma2020similarity}. While it does not achieve the top rank in $\rho$ and $\tau$, it performs competitively across all metrics. It is worth noting that CA-SUM~\cite{SummarizingACheat} trained five models with different $\delta$ values (0.5–0.9) for 400 epochs per fold, saving checkpoints at each epoch, resulting in 2000 checkpoints per fold. The best-performing checkpoint was then selected based on maximizing $\rho$ and $\tau$, which may have contributed to its high scores. Additionally, although another method achieves high correlation scores, it reports much lower F-scores, likely due to the lack of a temporal modeling component. This can lead to frame selections that are individually relevant but fail to form coherent summaries.
}
\subsubsection{Computational Efficiency and Stability Analysis}
{For a stability comparison across different methods, we report the standard deviation (SD) of F-scores in Table~\ref{tb:soc}. These values are derived by repeating the 5-fold experiment five times (25 total runs), computing the SD separately for each fold, and then averaging these SD values across all folds. The SumMe benchmark, with its smaller dataset and higher noise levels, serves as a more stringent test for assessing methods' stability.}

{In Table~\ref{tb:soc}, the GAN-based approaches~\cite{apostolidis2020unsupervised, apostolidis2020ac, yaliniz2021using} exhibit significantly higher SD values compared to our method, which can be attributed to the inherent instability of GAN training and its sensitivity to small changes in adversarial learning~\cite{arjovsky2017towards}. To gain a better understanding of this instability, we compared the training loss curves of our method with those of the latest GAN-based SOTA approaches~\cite{apostolidis2020unsupervised, apostolidis2020ac}, which made their source code publicly available, ensuring reproducibility. The comparison was performed using the fifth split of TVSum, where all models performed optimally. By selecting this split, we minimized data noise, allowing for a more fair comparison of the models' intrinsic capabilities. The results of this evaluation are shown in Fig.~\ref{fg:loss}. }

{SUM-GAN-AAE employs three distinct losses during training, as described in~\cite{apostolidis2020unsupervised}, while AC-SUM-GAN uses six different losses, as outlined in~\cite{apostolidis2020ac}. In Fig.~\ref{fg:loss}.a, three losses work in conjunction with each other, showing stagnant behavior and unstable convergence. Fig.~\ref{fg:loss}.b shows that in AC-SUM-GAN some losses exhibit oscillations, sudden spikes, and premature convergence.} {However, In Fig.~\ref{fg:loss}.c (our proposed method) the losses from two separate training stages, exhibit smooth and stable convergence.}

{To quantify stability, we analyzed the F-score performance curves over 100 epochs from all five TVSum splits, providing a standardized comparison not possible with the models' disparate loss functions. The analysis in Table~\ref{tb:qv} uses several key metrics. TVn measures the overall "jerkiness," MS quantifies consistent improvement, and RAUC calculates the cumulative "missed opportunity." Focusing on the critical final training phase, TV captures performance fluctuations while TS determines the trend over the last 20 epochs to see if the model has successfully plateaued~\cite{barinov2023automatic}. The results show that the TR-SUM method decisively outperforms the baselines across all metrics, confirming its ability to avoid the erratic training behaviors characteristic of GAN-based approaches.}

\begin{table}
 \centering
\begin{threeparttable}
  \centering
  \caption{{A quantitative comparison of model stability on F-score.}}
    \begin{tabular}{c|cccccc}
    \toprule
          & Avg (SD) $\Delta_{F}$ & TS    & TV    & RAUC  & MS    & TVn \\
    \midrule
    ~\cite{apostolidis2020unsupervised} & 1.4(0.5) & 0.0131 & 0.04  & 0.22  & 0.05  & 3.03  \\
    ~\cite{apostolidis2020ac} & 0.6(0.9) & 0.0112 & 0.11  & 0.54  & -0.02 & 7.65    \\
    TR-SUM & \textbf{1.5(0.3)} & \textbf{0.0004} & \textbf{0.03} & \textbf{0.12} & \textbf{0.10} & \textbf{1.93} \\
    \bottomrule
    \end{tabular}%
  \label{tb:qv}%
  \begin{tablenotes}
    \item TS: Tail Slope, TV: Tail Volatility, RAUC: Regret AUC, MS: Monotonicity Score, TVn: Normalized Total Variation
\end{tablenotes}
\end{threeparttable}
\end{table}%

% {In addition to recording training losses (Fig.~\ref{fg:loss}), {we measured the F-score at the 1st, 50th, and 100th epochs. SUM-GAN-AEE achieved scores of $59.8, 61.4, 61.2$, while AC-SUM-GAN obtained $61.3, 62.8, 61.8$. Our method delivered F-scores of $63.6, 65.6, 66.1$}. Our approach achieves a high F-score as early as the 1st epoch, benefiting from initializing the summarizer’s weights with the generator’s. It also demonstrates the greatest improvement after 100 epochs, with steady gains throughout training. Other models, however, show a decrease in the F-score after extended training, suggesting their optimization does not consistently enhance summarization quality
}

{According to Table~\ref{tb:soc}, RS-SUM is the only method with a lower SD value than TR-SUM. However, RS-SUM's stability comes at a high computational price: unlike TR-SUM's single-pass architecture, RS-SUM requires an iterative mechanism to refine predictions. This efficiency trade-off is quantified in Table~\ref{tb:time}. For a 128-frame sequence, TR-SUM is 310 times faster and 20 times less computationally expensive in terms of MACs. While TR-SUM has a higher parameter count and GPU memory footprint, these can be reduced by using a linear compression layer~\cite{myicip2023sum} or by reducing $l$.}
\begin{table}[t] 
\setlength{\tabcolsep}{2pt}
  \centering
  \caption{{Computational cost of various unsupervised methods.}}
\begin{tabular}{c|cccc}
\toprule
Method & Time (s) & FLOPs & Params (M) & GPU Mem (MB) \\
\midrule
DR-DSN~\cite{zhou2018deep} & 0.063 & 3.4E+08 & 2.6   & 48.9 \\
{CSNet}~\cite{jung2019discriminative} & 0.187 & 1.0E+09 & 6.3   & 47.9 \\
{SUM-GAN-sl}~\cite{apostolidis2019stepwise} & 0.07  & 1.6E+09 & 12.6  & 141.2 \\
SUM-GAN-AAE~\cite{apostolidis2020unsupervised} & 0.07  & 1.6E+09 & 12.6  & 141.2 \\
AC-SUM-GAN~\cite{apostolidis2020ac} & 0.201 & 3.5E+08 & 21.3  & 193.2 \\
CA-SUM~\cite{SummarizingACheat} & 0.007 & 6.7E+08 & 5.3   & 32.9 \\
{SegSum}~\cite{vo2025integrate} & 0.161 & 1.6E+06 & 5.3   & 28.84 \\
RS-SUM~\cite{myicip2023sum} & 2.718 & 9.7E+10 & 4.2   & 40.2 \\
TR-SUM & 0.009 & 4.8E+09 & 37.8  & 263.5 \\
\bottomrule
\end{tabular}%

  \label{tb:time}%
\end{table}%

\subsection{Ablation Study}
In this section, we perform an ablation study to investigate the impact of various parameters on the performance of the proposed model. We categorize our study into four subsections: Video decomposition, model configuration, self-supervised training, and summarizer training parameters. Table~\ref{tb:ablP} presents these parameters, and the sections they correspond to, and provides a brief description of each section.

\begin{table}[t] 
  \centering
  \setlength{\tabcolsep}{4pt}

  \caption{Description of important parameters studied in this ablation.}
    \begin{tabular}{c|l}
    \toprule
    \textbf{Stage} & \textbf{Parameter} \\
    \midrule
    Video decomposition & $L$ Segment's length \\
    \midrule
    \multirow{2}[2]{*}{Model configuration}
      & $l$: number of multi-head attention layers \\
      & $h$: number of attention heads per each layer \\
    \midrule
    \multirow{3}[2]{*}{ Self-supervised training} & Frame masking method \\
      & $M_R$: Masking ratio \\
      & Reconstruction loss function \\
    \midrule
    \multirow{2}[2]{*}{ Summarizer training} & $\delta$: Regularization Factor \\
      & $\beta$: Regularization Factor Coefficient \\
    \bottomrule
    \end{tabular}%
  \label{tb:ablP}%
\end{table}%
{To establish a base model for this work, we conducted exhaustive search experiments on key parameters: $L$, $l$, $h$, and the masking method. We set $M_R$ and the reconstruction loss function to values determined in the previous work~\cite{myicip2023sum}, and fixed $\delta$ and $\beta$ to values in between selected numbers. This exhaustive search experiment prioritized achieving a higher F-score over $\tau$ and $\rho$. After determining the base model, we conducted subsequent ablation studies, adjusting one parameter at a time while keeping the baseline values unchanged.}
\subsubsection{Video decomposition}
\begin{figure}[t!]
   \begin{center}
  \includegraphics[width=0.49\textwidth]{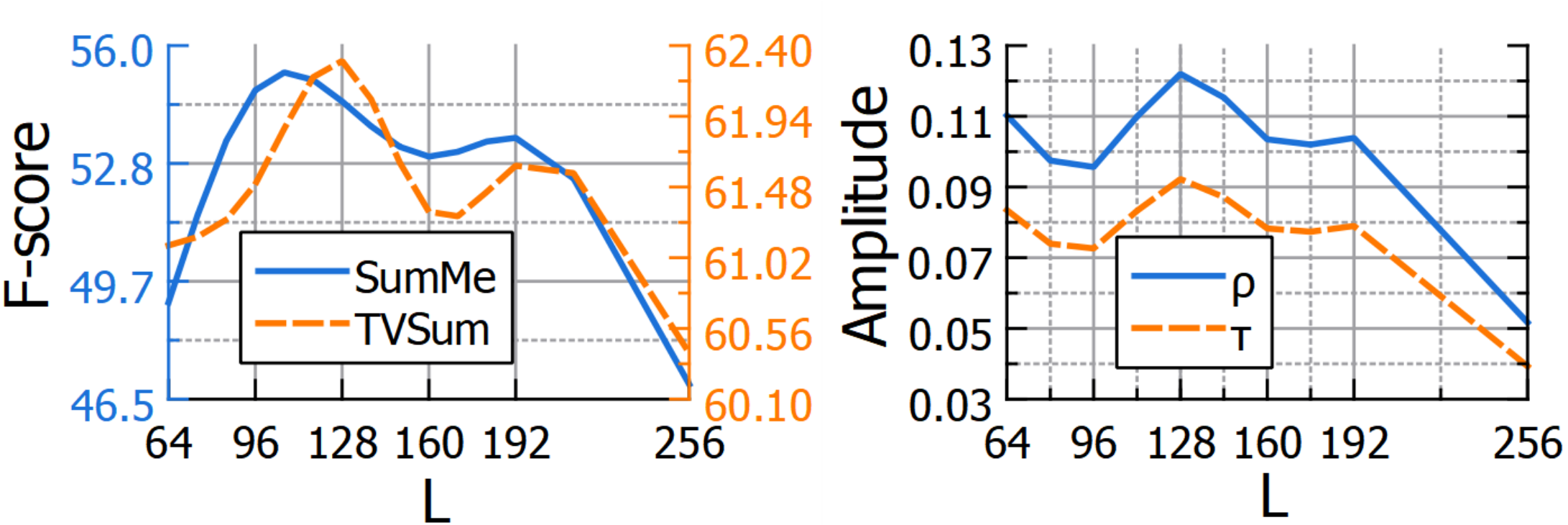} \caption{Effect of $L$ on F-score, $\rho$, and $\tau$.}\label{fg:L}
  \end{center}

\end{figure}
This section focuses solely on the parameter $L$, representing the length of the segments. The impact of changing $L$ on the model's F-score is illustrated in Fig.~\ref{fg:L}. Selecting extremely high or low values for $L$ results in a F-score decline.
The figure suggests that the optimal point for attaining a satisfactory F-score on both datasets is around $L=128$. Regarding $\rho$ and $\tau$, an increase in sequence length appears to reduce these values, similar to the results we obtained for the F-score.  Interestingly, $\rho$ and $\tau$ also peak at $L=128$, further reinforcing the significance of that value.

\subsubsection{Model configuration parameters}

The two parameters investigated in this section are $l$ (the number of layers) and $h$ (the number of attention heads) of the transformer model. 
Table~\ref{tb:lh} shows the impact of these parameters on the performance of the proposed model ({The bold values in each column indicate the highest score achieved for that metric}). As seen from these results, increasing $l$ from 1 to 3 enhances the model's F-score. However, for $l > $3, no specific trend is observed. While a single-layer transformer may not be sufficient to capture complex relationships, employing too many layers increases the risk of overfitting during the first training phase. Regarding $h$, the results suggest that an increase in $h$ enhances the $\rho$ and $\tau$. With $l$ set to the optimal value of 3 and $h$ to 8, we observed the highest performance, indicating a synergistic effect between these parameters.
\begin{table}[t] 
  \centering
  \caption{The effect of the number of attention heads and layers on the summary evaluation metrics.}

\begin{tabular}{c|c|c|c|c|c}
\toprule
\multicolumn{2}{c|}{Dataset} & \multicolumn{1}{l|}{SumMe} & \multicolumn{3}{c}{TVSum} \\
\midrule
$l$ & $h$ & Fscore & Fscore & $\tau$ & $\rho$ \\
\midrule
1 & 1 & 50.2 & 61.2 & 0.051 & 0.067 \\
1 & 4 & {53.6} & {61.6} & 0.070 & 0.093 \\
1 & 8 & 49.7 & 61.6 & {0.074} & {0.098} \\
1 & 16 & 51.2 & 59.9 & 0.027 & 0.035 \\
\midrule
3 & 1 & 53.1 & 61.4 & 0.037 & 0.049 \\
3 & 4 & 53.4 & 61.8 & 0.051 & 0.067 \\
3 & 8 & \textbf{54.5} & \textbf{62.3} & 0.092 & 0.122 \\
3 & 16 & 53.0 & 61.5 & \textbf{0.105} & \textbf{0.138} \\
\midrule
6 & 1 & 54.0 & 61.7 & 0.053 & 0.069 \\
6 & 4 & {54.3} & {61.7} & 0.062 & 0.081 \\
6 & 8 & 52.8 & 61.2 & 0.063 & 0.084 \\
6 & 16 & 52.4 & 61.1 & {0.069} & {0.090} \\
\bottomrule
\end{tabular}%

  \label{tb:lh}%
\end{table}%

\subsubsection{Self-supervised training parameters}
A key factor to consider is the effect of the dynamic masking method on the quality of generated summaries. As an alternative to the dynamic shot masking method proposed here, one could consider randomly masking a selection of frames during self-supervised training or adopting the method proposed in RS-SUM. The latter utilizes a fixed-size window masking scheme within each shot, in contrast to our method, which employs windows with dynamic lengths that expand or shrink based on the shot length. Table~\ref{tb:mm} presents the results of this comparison. In this table, $W_s$ denotes the window size used in the fixed window masking method. From these results, the dynamic masking method produces the best outcomes. {As shown in this table, increasing $W_s$ beyond 15 has minimal impact on performance. This is because the pre-processing and processing steps downsample shots into subsequences with few frames, causing larger window sizes to mask entire shots.}
\begin{table}
\setlength{\tabcolsep}{2pt}

  \centering
  \caption{{The effect of the masking method on the summary evaluation.}}
    
\begin{tabular}{l|c|c|c|c}
\toprule
Dataset & SumMe & \multicolumn{3}{c}{TVSum} \\
\midrule
Masking method & F-score & F-score & $\tau$ & $\rho$ \\
\midrule
Random Masking & 52.9 & 61.9 & 0.088 & 0.117 \\
Fixed window masking ($W_s$=3) & 53.8 & 61.6 & 0.060 & 0.079 \\
Fixed window masking ($W_s$=7) & 53.0 & 61.2 & 0.072 & 0.096 \\
Fixed window masking ($W_s$=11) & 53.4 & 61.6 & 0.068 & 0.090 \\
Fixed window masking ($W_s$=15) & 53.0 & 61.1 & 0.066 & 0.087 \\
Fixed window masking ($W_s$=31) & 53.4 & 61.0 & 0.064 & 0.084 \\
Fixed window masking ($W_s$=63) & 53.0 & 61.3 & 0.068 & 0.090 \\
\textbf{Dynamic masking} & \textbf{54.5} & \textbf{62.3} & \textbf{0.092} & \textbf{0.122} \\
\bottomrule
\end{tabular}%

  \label{tb:mm}%
\end{table}%
\begin{figure}[t]
   \begin{center}
  \includegraphics[width=0.35\textwidth]{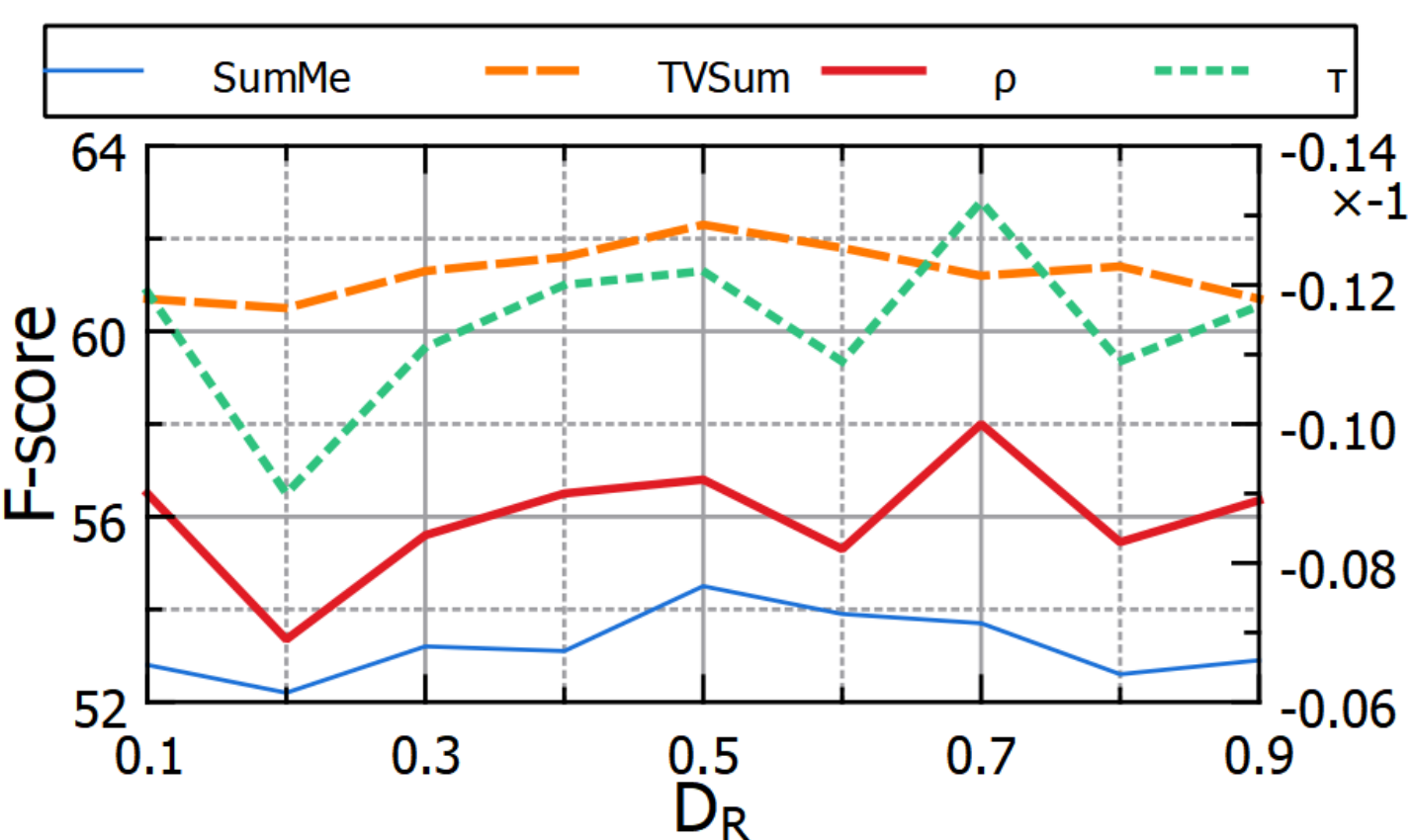} \caption{{The effect of using different $D_R$ values on the metric.}}\label{fg:dr}
  \end{center}

\end{figure}
\begin{figure}[t!]
   \begin{center}
  \includegraphics[width=0.45\textwidth]{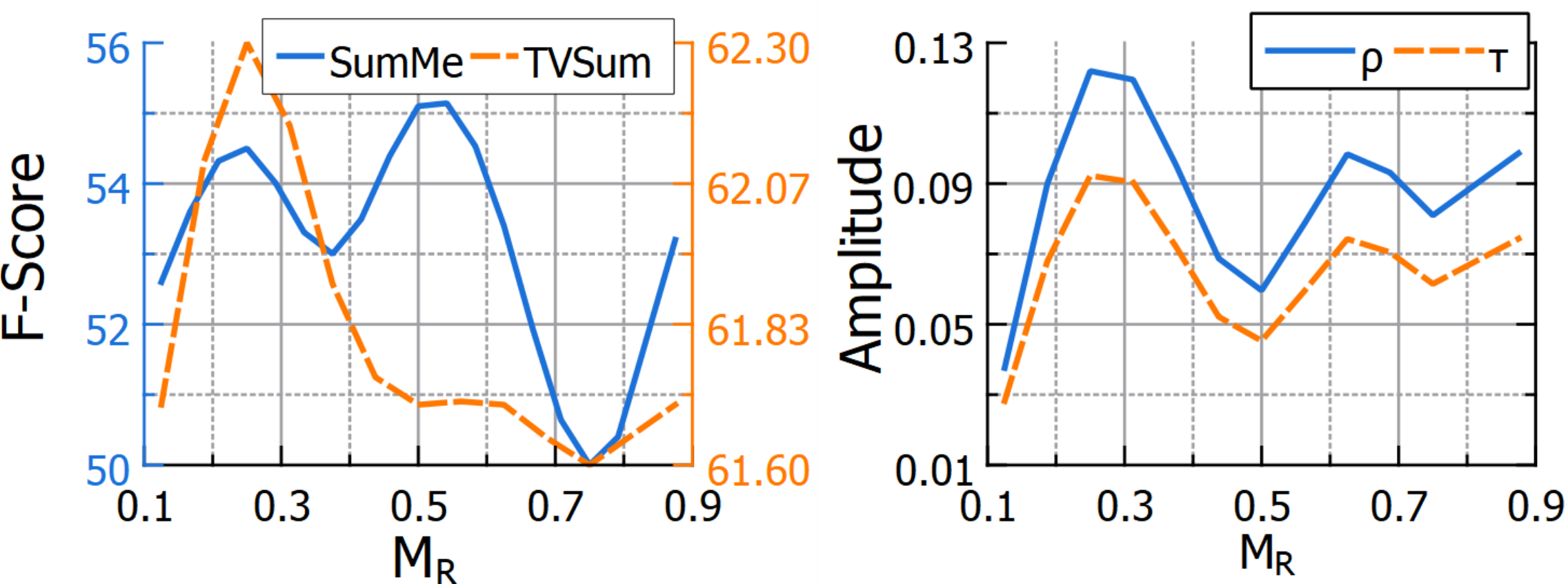} \caption{The effect of Masking ratio on F-score, $\rho$, and $\tau$.}\label{fg:mr}
  \end{center}

\end{figure}
{Another important factor is the masking ratio $ M_R $, defined as the proportion of masked frames to the total length of each sub-sequence. As shown in Fig.~\ref{fg:mr}, model performance peaks when $ M_R = 0.25 $ on both benchmark datasets. This ratio provides the generator with sufficient context to learn temporal dependencies while maintaining a meaningful reconstruction challenge. When $ M_R $ is too low, the generator fails to generalize to inputs with large missing segments, which are typical during training for highly compressed summaries. Conversely, if $ M_R $ is too high, the context becomes too sparse to support effective learning.}

\begin{figure}[t]
   \begin{center}
  \includegraphics[width=0.45\textwidth]{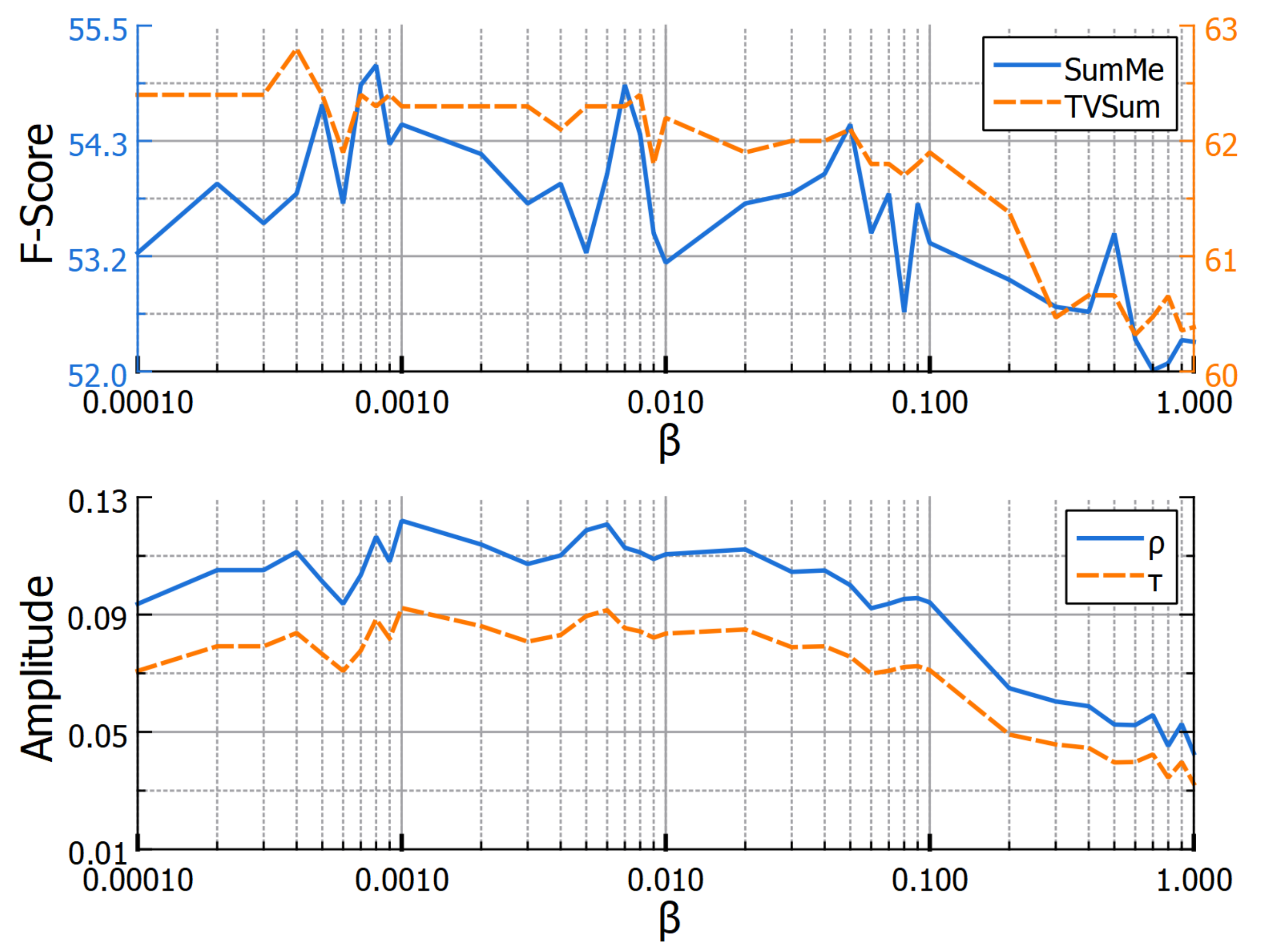} \caption{The effect of $\beta$ on F-score, $\rho$, and $\tau$.}\label{fg:beta}
  \end{center}

\end{figure}
\begin{table}[t!] 
\setlength{\tabcolsep}{3pt}
  \centering
  \caption{The effect of the reconstruction loss function.}

\begin{tabular}{l|c|c|c|c}
\toprule
Dataset & SumMe & \multicolumn{3}{c}{TVSum} \\
\midrule
Reconstruction Loss Function & F-score & F-score & $\tau$ & $\rho$ \\
\midrule
CE & 53.9 & 60.5 & 0.032 & 0.042 \\
L1 & 53.3 & 61.7 & 0.077 & 0.102 \\
MSE & 52.6 & 61.9 & 0.068 & 0.090 \\
MSE+CE & 54.0 & 61.5 & 0.057 & 0.076 \\
L1+CE & \textbf{54.5} & \textbf{62.3} & \textbf{0.092} & \textbf{0.122} \\
\bottomrule
\end{tabular}%

  \label{tb:loss}%
\end{table}%
\begin{figure}[ht]
   \begin{center}
  \includegraphics[width=0.49\textwidth]{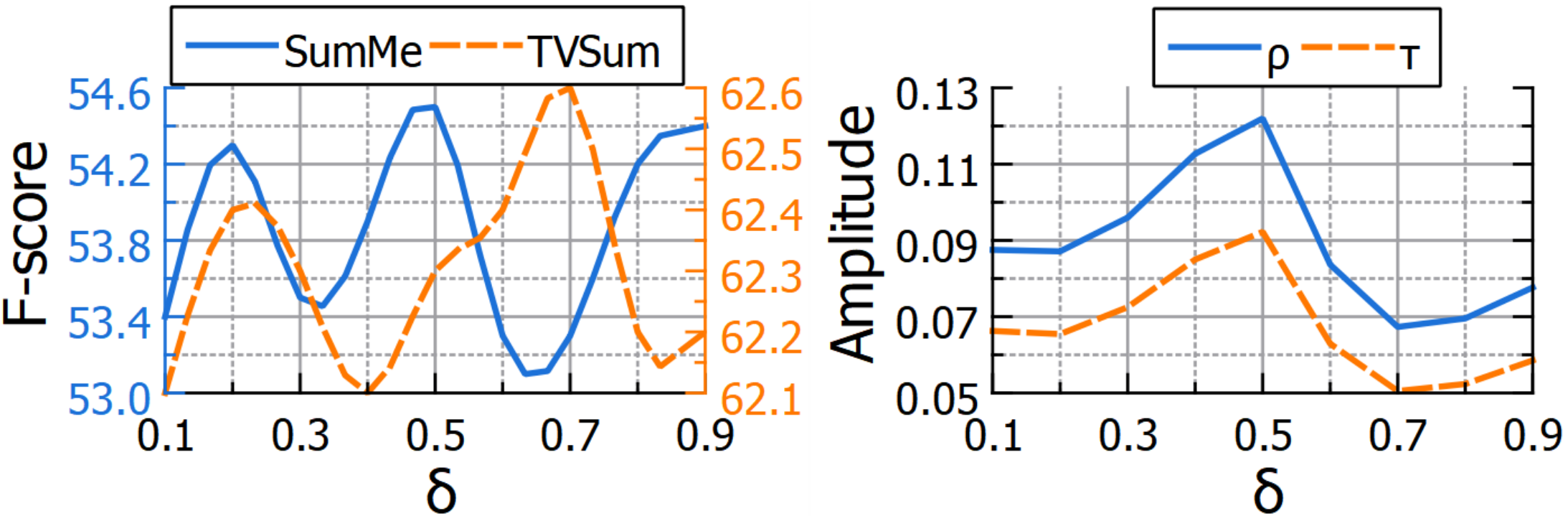} \caption{The effect of $\delta$ on F-score, $\rho$, and $\tau$.}\label{fg:delta}
  \end{center}

\end{figure}
\begin{figure}[h!]
   \begin{center}
  \includegraphics[width=0.45\textwidth]{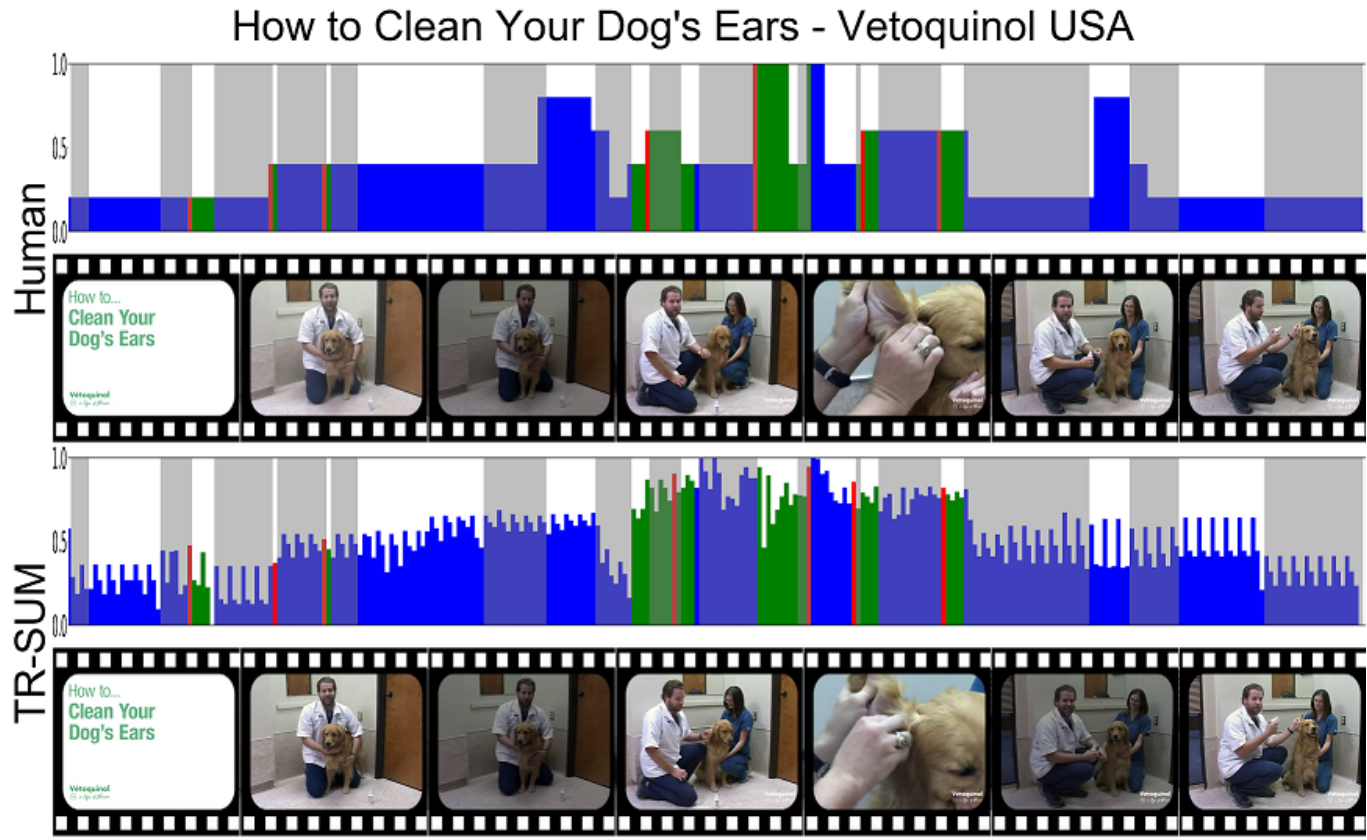} \caption{
Human-generated frame scores and summaries vs. our method's.
}\label{fig:vid1}
  \end{center}

\end{figure}

\begin{figure}[h!]
   \begin{center}
  \includegraphics[width=0.42\textwidth]{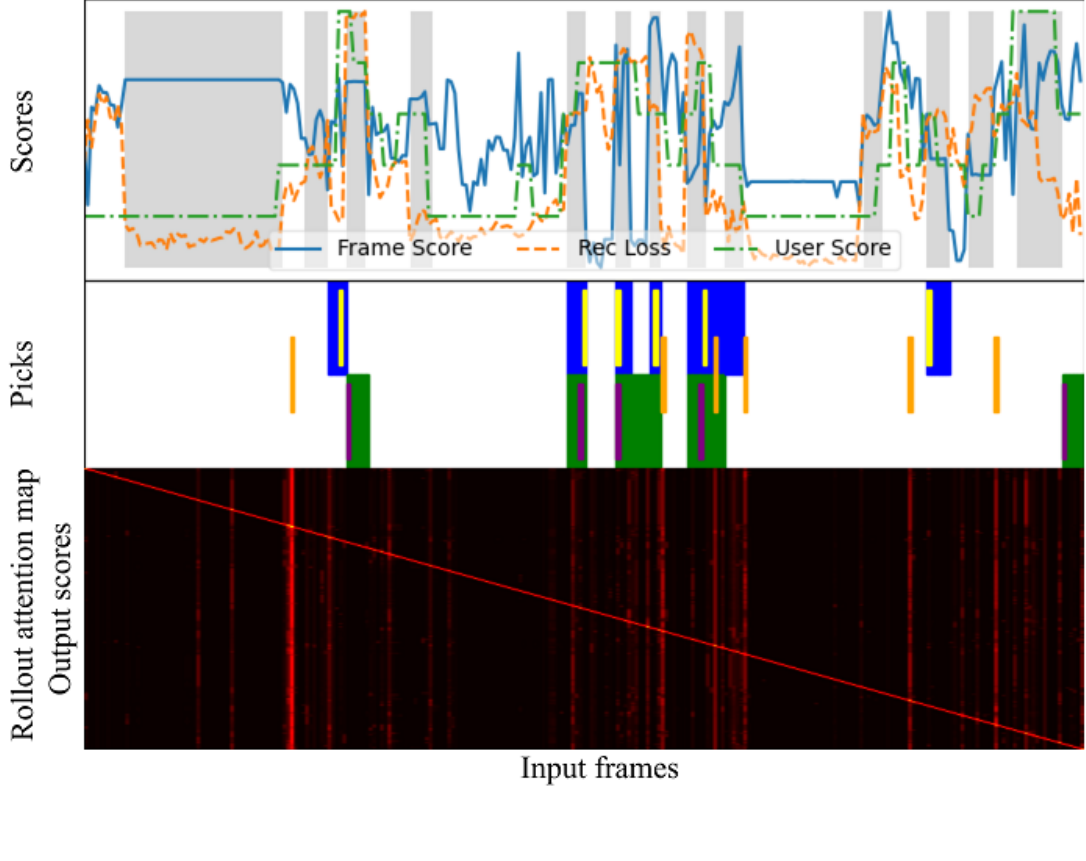} \caption{{
User scores, one-out reconstruction loss, and TR-SUM frame scores (top row); human vs. TR-SUM selected segments (green and blue, middle row); and rollout attention map (bottom row).}
}\label{fig:map}
  \end{center}

\end{figure}

The last parameter evaluated in this section is the reconstruction loss function defined in~\eqref{eq:ls}. In Table~\ref{tb:loss}, we show the effect of using alternative reconstruction loss functions such as MSE. MSE and L1 appear to perform similarly; however, the combination of L1 with CE outperforms all single combinations, including the combination of MSE with CE loss function. This underscores the superiority of the proposed combination in achieving better results.


\subsubsection{Summarizer's training parameters}

{
In this section, we examine the impact of two key training parameters, namely $\delta$ and $\beta$, as defined in~\eqref{eq:reg} and~\eqref{eq:sgd}. As $\beta$ increases, model performance tends to decline, since the loss becomes dominated by the regularization term, diminishing the influence of the reward signal. Conversely, when $\beta$ is too small, the reward dominates the loss and encourages the model to assign high scores to all frames. This leads to fewer frames being masked, making the generator's reconstruction task overly simple and degrading the learning signal. Therefore, we expect model performance to initially improve with increasing $\beta$, as the loss becomes better balanced, and then decline once regularization outweighs the reward. This trend is reflected in the metrics $\rho$ and $\tau$, as shown in Fig.~\ref{fg:beta}.

}

% {Additionally, the effect of $\delta$ on the F-score is illustrated in Fig.~\ref{fg:delta}. The influence of $\delta$ varies across different datasets. As shown in previous studies~\cite{yuan2019cycle,zhou2018deep,apostolidis2019stepwise,apostolidis2020unsupervised,apostolidis2020ac} and discussed in Section~\ref{sec:exp1}, a model typically achieves its peak performance at different $\delta$ values for each dataset.}
{Additionally, the effect of $\delta$ on the F-score is shown in Fig.~\ref{fg:delta}. While the F-score exhibits some volatility, the overall change in performance is relatively small. As discussed in Section~\ref{sec:exp1}, $\delta$ influences the average frame scores, which in turn affects how summaries are constructed, leading to variability in F-score across different splits. However, the evaluation metrics $\rho$ and $\tau$ are not affected by such change and show a clear peak at $\delta = 0.5$. }

% The reason $\delta$ impacts each dataset differently could be attributed to the summary generation stage, where the relationship between the distribution of scores and shot length determines the shots that will be selected. Given that datasets have different distributions of shot lengths, different distributions of frame scores are required to achieve a better F-score. Since $\delta$ directly influences the mean or center of the distribution, and the two datasets have quite different shot formats, different patterns of F-score against $\delta$ are observed on each dataset.}

{The ablation results indicate that hyperparameters directly affecting the generator training stage have the greatest impact on final performance, with $L$ being the most influential. $L$ controls both the context captured in each segment and the total number of training samples.}

\subsection{Qualitative visual analysis of the generated summaries}

\begin{figure}[t!]
  \centering
  \includegraphics[width=0.49\textwidth]{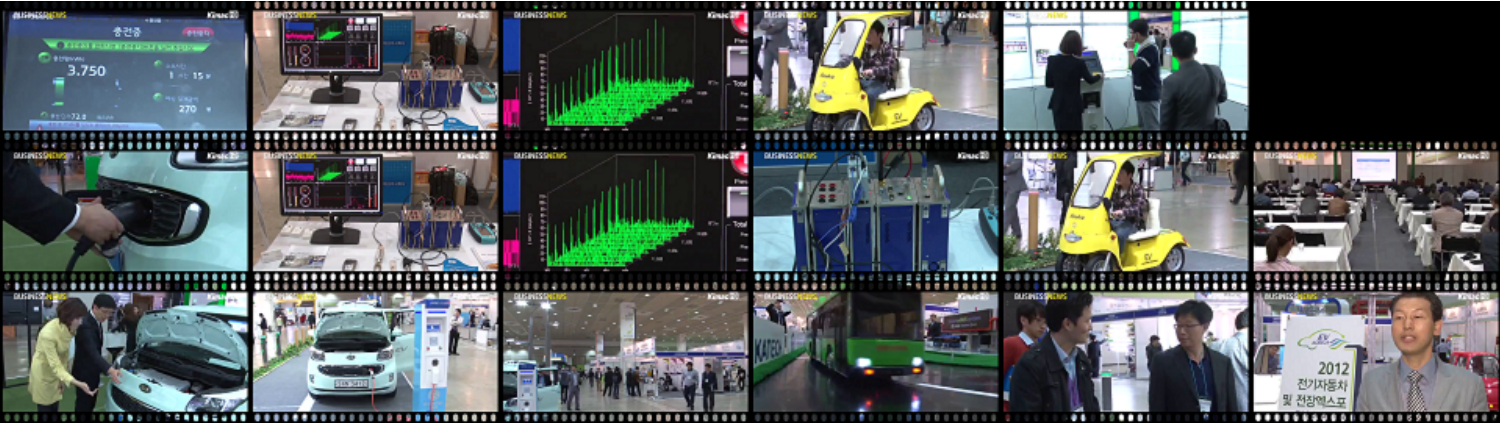} \caption{{Top-scoring summary frames by (top row) human annotator, (middle row) TR-SUM, and (bottom row) frames with highest attention.}}\label{fig:vid2}
  % \end{center}

\end{figure}
% {In this section, we present visual samples to demonstrate the effectiveness of our method and explore the relationships between the generated frame scores, our reward function, and the video semantics. Fig.~\ref{fig:vid1} illustrates the annotations provided by both our method and human observers. This figure is divided into two sections: ``Human'' and ``TR-SUM''. Each section features bar plots that display the normalized frame scores in blue, with alternating white and gray background colors indicating the beginning or end of a new shot. Segments highlighted in green represent the shots selected for the summary, while the highest-scoring frame within each green segment is marked in red. These frames are arranged horizontally and displayed in the second row beneath the bar plot. In the "Human" section, the annotation with the highest F-score compared to the other annotations is shown.

In this section, we present visual samples to demonstrate our method's effectiveness by exploring how the generated scores, our reward function, and video semantics are related. Fig.~\ref{fig:vid1} compares annotations from our method and human observers. The figure is split into ``Human'' and ``TR-SUM'' sections. Each section has bar plots showing normalized frame scores in blue, with alternating white and gray backgrounds to indicate new shots. Green highlights show selected summary shots, and the highest-scoring frame within each is marked in red and displayed below the plot. For the ``Human'' section, we show the annotation with the highest F-score compared to other annotations.

The ``TR-SUM'' section presents the frame scores and summary generated by our algorithm. Fig~\ref{fig:vid1}
reveals significant similarities between the key frames selected by the human annotator and those chosen by our model. To further examine the relationship between video semantics and frame score generation, we turn to Fig.~\ref{fig:map}. The top row displays a comparison between user-assigned frame scores, TR-SUM frame scores, and one-out reconstruction loss. The one-out reconstruction loss for each frame is calculated by replacing it with the mask token, and reconstructing it using the generator while keeping the remaining frames unchanged. The second row highlights segments selected by the human annotator in green, with the highest-scoring frame within each segment indicated by a purple bar. The sections chosen by the algorithm are marked with blue bars, and the highest-scoring frame in each segment is shown in yellow. Finally, the orange bars indicate the frames with high attention scores. The last row presents the summarizer's rollout attention map between the input frames and the output score. Frames marked in purple, yellow, and orange are shown in Fig.~\ref{fig:vid2} in the top row, middle row, and bottom row, respectively.}

{A visual comparison between the one-out reconstruction loss and user-assigned frame importance scores reveals a strong correlation. The positive correlation shown in the figure demonstrates that frames critical for reconstruction are also those that humans perceive as important, supporting the validity of our reward function.} By examining the relationship between the rollout attention map and the one-out reconstruction loss, we observe that key frames (i.e., frames receiving the highest attention) are mostly located at the start or end of certain shots and are associated with shifts in reconstruction loss. Further analysis of the rollout attention map reveals that the model assigns frame importance by assessing each frame’s relation to these key frames, which serve as representations of the video’s core narrative. Essentially, the model prioritizes frames that are both unique and contextually significant and connected, as they contribute most to minimizing reconstruction loss. Inadvertently, the model learns to assign high scores to frames from long shots, as these frames contribute significantly to the reconstruction of other frames within the same shots. However, due to the long shot length, these frames are ignored by the knapsack algorithm.
\section {Discussion and Future Work}\label{sec:disc}

{Our reward generation pipeline is, in principle, fully differentiable and could be paired with differentiable alternatives to Bernoulli sampling. In practice we retained a reinforcement-learning setup because training with soft sampling reduced reconstruction fidelity and differentiable approximations produced unstable gradients and reconstruction losses that did not reliably track true frame importance. Keeping the current reward formulation also facilitates integration with complementary objectives such as diversity, representativeness~\cite{zhou2018deep}, shot semantics~\cite{yuan2022unsupervised}, or multimodal rewards~\cite{barbakos2025unsupervised}.}

{Our method has several known limitations because its indirect optimization process can cause the model to prioritize visual fidelity over subtler semantic cues. The concrete failure cases include: (a) redundancy in repetitive scenes, where visually similar frames are scored highly, producing less diverse summaries; (b) underselection of brief events, as short but important moments are missed due to their minimal impact on reconstruction error; (c) a preference for transition frames, with abrupt changes receiving high scores even though people tend to exclude them; and (d) modality and domain gaps, since the visual-only input misses audio cues and performance on domains like surveillance footage has not been validated.}

{Promising directions to address these boundaries include designing differentiable reconstruction-based rewards that combine fidelity with semantic objectives, as well as fusing contrastive and temporal modeling to improve diversity and narrative coherence. Another important direction is to enrich the reward with multimodal and structural cues such as audio, shot semantics, and explicit objectives for diversity and representativeness, which can mitigate redundancy, missed brief events, and the modality gap. Finally, incorporating richer embeddings such as CLIP or spatiotemporal 3D CNN features, together with domain adaptation and updated evaluation protocols, will support robustness across challenging video domains and enable more realistic assessments of progress.}
\section{Conclusion}\label{sec:con}

This paper introduced an unsupervised approach to video summarization using reinforcement learning with a learnable reward pipeline. Unlike previous methods that used manual reward functions, our approach employs a trained video generator to reconstruct masked frames, deriving rewards from the similarity between reconstructed and original videos. The underlying principle is that informative summaries will enable better video reconstruction. The video generator, trained through self-supervised learning, serves as pre-training for the summarizer, which independently generates frame scores during inference. Experimental results on TVSum and SumMe datasets demonstrate our method's superiority, achieving F-scores of 62.3 and 54.5 respectively, highlighting its potential for producing high-quality video summaries.

\bibliographystyle{IEEEtran}
\bibliography{bare_jrnl_new_sample4.bib}

@inproceedings{mahasseni2017unsupervised,
  title={Unsupervised video summarization with adversarial lstm networks},
  author={Mahasseni, Behrooz and Lam, Michael and Todorovic, Sinisa},
  booktitle={Proc. IEEE Conf. Comput. Vis. Pattern Recognit.},
  pages={202--211},
  year={2017}
}

@inproceedings{apostolidis2019stepwise,
  title={A stepwise, label-based approach for improving the adversarial training in unsupervised video summarization},
  author={Apostolidis, Evlampios and Metsai, Alexandros I and Adamantidou, Eleni and Mezaris, Vasileios and Patras, Ioannis},
  booktitle={Proc. 1st Int. Workshop AI Smart TV Content Prod. Access Del.},
  pages={17--25},
  year={2019}
}

@inproceedings{carreira2017quo,
  title={Quo vadis, action recognition? a new model and the kinetics dataset},
  author={Carreira, Joao and Zisserman, Andrew},
  booktitle={Proc. IEEE Conf. Comput. Vis. Pattern Recognit.},
  pages={6299--6308},
  year={2017}
}

@book{sutton2018reinforcement,
  title={Reinforcement learning: An introduction},
  author={Sutton, Richard S and Barto, Andrew G},
  year={2018},
  publisher={MIT press}
}

@article{sewak2019policy,
  title={Policy-based reinforcement learning approaches: Stochastic policy gradient and the REINFORCE algorithm},
  author={Sewak, Mohit and Sewak, Mohit},
  journal={Deep Reinforcement Learn.: Frontiers Artif. Intell.},
  pages={127--140},
  year={2019},
  publisher={Springer}
}

@article{williams1992simple,
  title={Simple statistical gradient-following algorithms for connectionist reinforcement learning},
  author={Williams, Ronald J},
  journal={Machine Learn.},
  volume={8},
  pages={229--256},
  year={1992},
  publisher={Springer}
}

@Article{yuan2019cycle,
  author    = {Yuan, Li and Tay, Francis Eng Hock and Li, Ping and Feng, Jiashi},
  title     = {Unsupervised video summarization with cycle-consistent adversarial LSTM networks},
  number    = {10},
  pages     = {2711--2722},
  volume    = {22},
  fjournal  = {IEEE Transactions on Multimedia},
  journal   = {IEEE Trans. Multimedia},
  publisher = {IEEE},
  year      = {2019},
}

@Article{apostolidis2020ac,
  author    = {Apostolidis, Evlampios and Adamantidou, Eleni and Metsai, Alexandros I and Mezaris, Vasileios and Patras, Ioannis},
  title     = {Ac-sum-gan: Connecting actor-critic and generative adversarial networks for unsupervised video summarization},
  number    = {8},
  pages     = {3278--3292},
  volume    = {31},
  fjournal  = {IEEE Transactions on Circuits and Systems for Video Technology},
  journal   = {IEEE Trans. Circuits Syst. Video Technol.},
  publisher = {IEEE},
  year      = {2020},
}

@inproceedings{jung2019discriminative,
  title={Discriminative feature learning for unsupervised video summarization},
  author={Jung, Yunjae and Cho, Donghyeon and Kim, Dahun and Woo, Sanghyun and Kweon, In So},
  booktitle={Proc. AAAI Conf. Artif. Intell.},
  volume={33},
  number={01},
  pages={8537--8544},
  year={2019}
}

@inproceedings{apostolidis2020unsupervised,
  title={Unsupervised video summarization via attention-driven adversarial learning},
  author={Apostolidis, Evlampios and Adamantidou, Eleni and Metsai, Alexandros I and Mezaris, Vasileios and Patras, Ioannis},
  booktitle={ICMM},
  pages={492--504},
  year={2020},
  organization={Springer}
}

@inproceedings{zhou2018deep,
  title={Deep reinforcement learning for unsupervised video summarization with diversity-representativeness reward},
  author={Zhou, Kaiyang and Qiao, Yu and Xiang, Tao},
  booktitle={Proc. AAAI Conf. Artif. Intell.},
  volume={32},
  number={1},
  year={2018}
}

@inproceedings{gonuguntla2019enhanced,
  title={Enhanced deep video summarization network},
  author={Gonuguntla, N and Mandal, B and Puhan, NB and others},
  year={2019},
  organization={BMVC}
}

@article{zhao2019property,
  title={Property-constrained dual learning for video summarization},
  author={Zhao, Bin and Li, Xuelong and Lu, Xiaoqiang},
  journal={IEEE Trans. Neural Netw. Learn. Syst.},
  volume={31},
  number={10},
  pages={3989--4000},
  year={2019},
  publisher={IEEE}
}

@inproceedings{jung2020global,
  title={Global-and-local relative position embedding for unsupervised video summarization},
  author={Jung, Yunjae and Cho, Donghyeon and Woo, Sanghyun and Kweon, In So},
  booktitle={ECCV},
  pages={167--183},
  year={2020},
  organization={Springer}
}

@inproceedings{liu2019learning,
  title={Learning hierarchical self-attention for video summarization},
  author={Liu, Yen-Ting and Li, Yu-Jhe and Yang, Fu-En and Chen, Shang-Fu and Wang, Yu-Chiang Frank},
  booktitle={2019 IEEE Int. Conf. Image Process. (ICIP)},
  pages={3377--3381},
  year={2019},
  organization={IEEE}
}

@inproceedings{he2019unsupervised,
  title={Unsupervised video summarization with attentive conditional generative adversarial networks},
  author={He, Xufeng and Hua, Yang and Song, Tao and Zhang, Zongpu and Xue, Zhengui and Ma, Ruhui and Robertson, Neil and Guan, Haibing},
  booktitle={Proc. 27th ACM Int. Conf. Multimedia},
  pages={2296--2304},
  year={2019}
}

@Article{yoon2021interp,
  author    = {Yoon, Ui-Nyoung and Hong, Myung-Duk and Jo, Geun-Sik},
  title     = {Interp-SUM: Unsupervised Video Summarization with Piecewise Linear Interpolation},
  number    = {13},
  pages     = {4562},
  volume    = {21},
  fjournal  = {Sensors},
  journal   = {Sensors-basel.},
  publisher = {Multidisciplinary Digital Publishing Institute},
  year      = {2021},
}

@inproceedings{li2018independently,
  title={Independently recurrent neural network (indrnn): Building a longer and deeper rnn},
  author={Li, Shuai and Li, Wanqing and Cook, Chris and Zhu, Ce and Gao, Yanbo},
  booktitle={Proc. IEEE Conf. Comput. Vis. Pattern Recognit.},
  pages={5457--5466},
  year={2018}
}

@article{vaswani2017attention,
  title={Attention is all you need},
  author={Vaswani, Ashish and Shazeer, Noam and Parmar, Niki and Uszkoreit, Jakob and Jones, Llion and Gomez, Aidan N and Kaiser, {\L}ukasz and Polosukhin, Illia},
  journal={Adv. Neural Inf. Process. Syst.},
  volume={30},
  year={2017}
}

@inproceedings{potapov2014category,
  title={Category-specific video summarization},
  author={Potapov, Danila and Douze, Matthijs and Harchaoui, Zaid and Schmid, Cordelia},
  booktitle={ECCV},
  pages={540--555},
  year={2014},
  organization={Springer}
}

@inproceedings{SummarizingACheat,
author = {Apostolidis, Evlampios and Balaouras, Georgios and Mezaris, Vasileios and Patras, Ioannis},
title = {Summarizing Videos Using Concentrated Attention and Considering the Uniqueness and Diversity of the Video Frames},
year = {2022},
isbn = {9781450392389},
publisher = {Association for Computing Machinery},
address = {New York, NY, USA},
pages = {407-415},
numpages = {9},
location = {Newark, NJ, USA},
series = {ICMR '22}
}

@inproceedings{TVSUM,
  title={Tvsum: Summarizing web videos using titles},
  author={Song, Yale and Vallmitjana, Jordi and Stent, Amanda and Jaimes, Alejandro},
  booktitle={Proc. IEEE Conf. Comput. Vis. Pattern Recognit.},
  pages={5179--5187},
  year={2015}
}

@inproceedings{SUMME,
  title={Creating summaries from user videos},
  author={Gygli, Michael and Grabner, Helmut and Riemenschneider, Hayko and Gool, Luc Van},
  booktitle={Eur. Conf. Comput. Vis.},
  pages={505--520},
  year={2014},
  organization={Springer}
}

@article{ovp,
  title={VSUMM: A mechanism designed to produce static video summaries and a novel evaluation method},
  author={De Avila, Sandra Eliza Fontes and Lopes, Ana Paula Brandao and da Luz Jr, Antonio and de Albuquerque Ara{\'u}jo, Arnaldo},
  journal={Pattern Recognit. Lett.},
  volume={32},
  number={1},
  pages={56--68},
  year={2011},
  publisher={Elsevier}
}

@Article{surv,
  author    = {Apostolidis, Evlampios and Adamantidou, Eleni and Metsai, Alexandros I and Mezaris, Vasileios and Patras, Ioannis},
  title     = {Video summarization using deep neural networks: A survey},
  number    = {11},
  pages     = {1838--1863},
  volume    = {109},
  fjournal  = {Proceedings of the IEEE},
  journal   = {Proc. IEEE},
  publisher = {IEEE},
  year      = {2021},
}

@article{kendall1945treatment,
  title={The treatment of ties in ranking problems},
  author={Kendall, Maurice G},
  journal={Biometrika},
  volume={33},
  number={3},
  pages={239--251},
  year={1945},
  publisher={JSTOR}
}

@book{zwillinger1999crc,
  title={CRC standard probability and statistics tables and formulae},
  author={Zwillinger, Daniel and Kokoska, Stephen},
  year={1999},
  publisher={Crc Press}
}

@inproceedings{zhang2016video,
  title={Video summarization with long short-term memory},
  author={Zhang, Ke and Chao, Wei-Lun and Sha, Fei and Grauman, Kristen},
  booktitle={Eur. Conf. Comput. Vis.},
  pages={766--782},
  year={2016},
  organization={Springer}
}

@inproceedings{szegedy2015going,
  title={Going deeper with convolutions},
  author={Szegedy, Christian and Liu, Wei and Jia, Yangqing and Sermanet, Pierre and Reed, Scott and Anguelov, Dragomir and Erhan, Dumitru and Vanhoucke, Vincent and Rabinovich, Andrew},
  booktitle={Proc. IEEE Conf. Comput. Vis. Pattern Recognit.},
  pages={1--9},
  year={2015}
}

@inproceedings{jadon2020unsupervised,
  title={Unsupervised video summarization framework using keyframe extraction and video skimming},
  author={Jadon, Shruti and Jasim, Mahmood},
  booktitle={2020 IEEE 5th Int. Conf. Comput. Commun. Autom. (ICCCA)},
  pages={140--145},
  year={2020},
  organization={IEEE}
}

@Article{zhu2022relational,
  author    = {Zhu, Wencheng and Han, Yucheng and Lu, Jiwen and Zhou, Jie},
  title     = {Relational Reasoning Over Spatial-Temporal Graphs for Video Summarization},
  pages     = {3017--3031},
  volume    = {31},
  fjournal  = {IEEE Transactions on Image Processing},
  journal   = {IEEE Trans. Image Process.},
  publisher = {IEEE},
  year      = {2022},
}

@inproceedings{ghauri2021supervised,
  title={Supervised video summarization via multiple feature sets with parallel attention},
  author={Ghauri, Junaid Ahmed and Hakimov, Sherzod and Ewerth, Ralph},
  booktitle={2021 IEEE Int. Conf. Multimedia Expo (ICME)},
  pages={1--6s},
  year={2021},
  organization={IEEE}
}

@inproceedings{apostolidis2021combining,
  title={Combining global and local attention with positional encoding for video summarization},
  author={Apostolidis, Evlampios and Balaouras, Georgios and Mezaris, Vasileios and Patras, Ioannis},
  booktitle={2021 IEEE Int. Symp. Multimedia (ISM)},
  pages={226--234},
  year={2021},
  organization={IEEE}
}

@article{myicip2023sum,
  author={Abbasi, Mehryar and Saeedi, Parvaneh},
  booktitle={2023 IEEE Int. Conf. Image Process. (ICIP)}, 
  title={Adopting Self-Supervised Learning into Unsupervised Video Summarization through Restorative Score.}, 
  year={2023},
  volume={},
  number={},
  pages={425-429},
  doi={10.1109/ICIP49359.2023.10222350}}

@inproceedings{otani2019rethinking,
  title={Rethinking the evaluation of video summaries},
  author={Otani, Mayu and Nakashima, Yuta and Rahtu, Esa and Heikkila, Janne},
  booktitle={Proc. IEEE/CVF Conf. Comput. Vis. Pattern Recognit.},
  pages={7596--7604},
  year={2019}
}

@article{ramachandran2017searching,
  title={Searching for activation functions},
  author={Ramachandran, Prajit and Zoph, Barret and Le, Quoc V},
  journal={arXiv preprint arXiv:1710.05941},
  year={2017}
}

@Article{springer,
  author    = {Tiwari, Vasudha and Bhatnagar, Charul},
  title     = {A survey of recent work on video summarization: approaches and techniques},
  number    = {18},
  pages     = {27187--27221},
  volume    = {80},
  fjournal  = {Multimedia Tools and Applications},
  journal   = {Multimed. Tools Appl.},
  publisher = {Springer},
  year      = {2021},
}

@article{deepai,
  title={Video Summarization Overview},
  author={Otani, Mayu and Song, Yale and Wang, Yang and others},
  journal={Found. Trends Comput. Graph. Vis.},
  volume={13},
  number={4},
  pages={284--335},
  year={2022},
  publisher={Now Publishers, Inc.}
}

@article{frontiers,
  title={From video summarization to real time video summarization in smart cities and beyond: A survey},
  author={Shambharkar, Prashant Giridhar and Goel, Ruchi},
  journal={Front. Big Data},
  volume={5},
  pages={1106776},
  year={2023},
  publisher={Frontiers}
}

@article{li2023unsupervised,
  title={Unsupervised Video Summarization},
  author={Li, Hanqing and Klabjan, Diego and Utke, Jean},
  journal={arXiv preprint arXiv:2311.03745},
  year={2023}
}

@article{arjovsky2017towards,
  title={Towards principled methods for training generative adversarial networks},
  author={Arjovsky, Martin and Bottou, L{\'e}on},
  journal={arXiv preprint arXiv:1701.04862},
  year={2017}
}

@article{surg,
  title={Surgical video summarization: Multifarious uses, summarization process and ad-hoc coordination},
  author={Avellino, Ignacio and Nozari, Sheida and Canlorbe, Geoffroy and Jansen, Yvonne},
  journal={Proc. ACM Hum.-Comput. Interact.},
  volume={5},
  number={CSCW1},
  pages={1--23},
  year={2021},
  publisher={ACM New York, NY, USA}
}

@inproceedings{gygli2016,
  title={Video2gif: Automatic generation of animated gifs from video},
  author={Gygli, Michael and Song, Yale and Cao, Liangliang},
  booktitle={IEEE Conf. Comput. Vis. Pattern Recognit.},
  pages={1001--1009},
  year={2016}
}

@Article{trafic,
  author    = {Sabha, Ambreen and Selwal, Arvind},
  title     = {Towards machine vision-based video analysis in smart cities: a survey, framework, applications and open issues},
  pages     = {1--52},
  fjournal  = {Multimedia Tools and Applications},
  journal   = {Multimed. Tools Appl.},
  publisher = {Springer},
  year      = {2023},
}

@Article{gupta2023,
  author    = {Gupta, Deeksha and Sharma, Akashdeep},
  title     = {A comprehensive study of automatic video summarization techniques},
  number    = {10},
  pages     = {11473--11633},
  volume    = {56},
  fjournal  = {Artificial Intelligence Review},
  journal   = {Artif. Intell. Rev.},
  publisher = {Springer},
  year      = {2023},
}

@inproceedings{wang2024m3sum,
  title={M3sum: A Novel Unsupervised Language-Guided Video Summarization},
  author={Wang, Hongru and Zhou, Baohang and Zhang, Zhengkun and Du, Yiming and Ho, David and Wong, Kam-Fai},
  booktitle={ICASSP 2024-2024 IEEE International Conference on Acoustics, Speech and Signal Processing (ICASSP)},
  pages={4140--4144},
  year={2024},
  organization={IEEE}
}

@Article{li2023progressive,
  author    = {Wang, Guolong and Wu, Xun and Yan, Junchi},
  title     = {Progressive reinforcement learning for video summarization},
  pages     = {119888},
  volume    = {655},
  fjournal  = {Information Sciences},
  journal   = {Inform. Sciences},
  publisher = {Elsevier},
  year      = {2024},
}

@article{huang2024aesthetic,
  title={An Aesthetic-driven Approach to Unsupervised Video Summarization},
  author={Huang, HongBen and Wu, ZaiQun and Pang, GuangYao and Xie, JieHang},
  journal={IEEE Access},
  year={2024},
  publisher={IEEE}
}

@article{narasimhan2021clip,
  title={Clip-it! language-guided video summarization},
  author={Narasimhan, Medhini and Rohrbach, Anna and Darrell, Trevor},
  journal={Advances in neural information processing systems},
  volume={34},
  pages={13988--14000},
  year={2021}
}

@article{sugihara2024language,
  title={Language-Guided Self-Supervised Video Summarization Using Text Semantic Matching Considering the Diversity of the Video},
  author={Sugihara, Tomoya and Masuda, Shuntaro and Xiao, Ling and Yamasaki, Toshihiko},
  journal={arXiv preprint arXiv:2405.08890},
  year={2024}
}

@Article{yu2024unsupervised,
  author    = {Yu, Qinghao and Yu, Hui and Sun, Ying and Ding, Derui and Jian, Muwei},
  title     = {Unsupervised Video Summarization Based on the Diffusion Model of Feature Fusion},
  fjournal  = {IEEE Transactions on Computational Social Systems},
  journal   = {IEEE Trans. Comput. Social Syst.},
  publisher = {IEEE},
  year      = {2024},
}

@Article{yaliniz2021using,
  author    = {Yaliniz, Gokhan and Ikizler-Cinbis, Nazli},
  title     = {Using independently recurrent networks for reinforcement learning based unsupervised video summarization},
  pages     = {17827--17847},
  volume    = {80},
  fjournal  = {Multimedia Tools and Applications},
  journal   = {Multimed. Tools Appl.},
  publisher = {Springer},
  year      = {2021},
}

@Article{greensmith2004variance,
  author   = {Greensmith, Evan and Bartlett, Peter L and Baxter, Jonathan},
  title    = {Variance Reduction Techniques for Gradient Estimates in Reinforcement Learning.},
  number   = {9},
  volume   = {5},
  fjournal = {Journal of Machine Learning Research},
  journal  = {J. Mach. Learn. Res.},
  year     = {2004},
}

@article{mao2018variance,
  title={Variance reduction for reinforcement learning in input-driven environments},
  author={Mao, Hongzi and Venkatakrishnan, Shaileshh Bojja and Schwarzkopf, Malte and Alizadeh, Mohammad},
  journal={arXiv:1807.02264},
  year={2018}
}

@Article{yuan2022unsupervised,
  author    = {Yuan, Ye and Zhang, Jiawan},
  title     = {Unsupervised video summarization via deep reinforcement learning with shot-level semantics},
  number    = {1},
  pages     = {445--456},
  volume    = {33},
  fjournal  = {IEEE Transactions on Circuits and Systems for Video Technology},
  journal   = {IEEE Trans. Circuits Syst. Video Technol.},
  publisher = {IEEE},
  year      = {2022},
}

@Article{xu2023self,
  author    = {Xu, Yifei and Li, Xiangshun and Pan, Litong and Sang, Weiguang and Wei, Pingping and Zhu, Li},
  title     = {Self-supervised adversarial video summarizer with context latent sequence learning},
  number    = {8},
  pages     = {4122--4136},
  volume    = {33},
  fjournal  = {IEEE Transactions on Circuits and Systems for Video Technology},
  journal   = {IEEE Trans. Circuits Syst. Video Technol.},
  publisher = {IEEE},
  year      = {2023},
}

@Article{ma2020similarity,
  author    = {Ma, Mingyang and Mei, Shaohui and Wan, Shuai and Wang, Zhiyong and Feng, David Dagan and Bennamoun, Mohammed},
  title     = {Similarity based block sparse subset selection for video summarization},
  number    = {10},
  pages     = {3967--3980},
  volume    = {31},
  fjournal  = {IEEE Transactions on Circuits and Systems for Video Technology},
  journal   = {IEEE Trans. Circuits Syst. Video Technol.},
  publisher = {IEEE},
  year      = {2020},
}

@article{barbakos2025unsupervised,
  title={Unsupervised Transcript-assisted Video Summarization and Highlight Detection},
  author={Barbakos, Spyros and Antoniadis, Charalampos and Potamianos, Gerasimos and Setti, Gianluca},
  journal={arXiv preprint arXiv:2505.23268},
  year={2025}
}

@inproceedings{pang2023contrastive,
  title={Contrastive losses are natural criteria for unsupervised video summarization},
  author={Pang, Zongshang and Nakashima, Yuta and Otani, Mayu and Nagahara, Hajime},
  booktitle={Proc. IEEE/CVF Winter Conf. Appl. Comput. Vis.},
  pages={2010--2019},
  year={2023}
}

@inproceedings{li2024unsupervised,
  title={Unsupervised Video Summarization via Iterative Training and Simplified GAN},
  author={Li, Hanqing and Klabjan, Diego and Utke, Jean},
  booktitle={Proceedings of the Asian Conference on Computer Vision},
  pages={1585--1601},
  year={2024}
}

@article{vo2025integrate,
  title={Integrate the temporal scheme for unsupervised video summarization via attention mechanism},
  author={Vo, Bang Q and Vo, Viet H},
  journal={IEEE Access},
  year={2025},
  publisher={IEEE}
}

@article{barinov2023automatic,
  title={Automatic evaluation of neural network training results},
  author={Barinov, Roman and Gai, Vasiliy and Kuznetsov, George and Golubenko, Vladimir},
  journal={Comput.},
  volume={12},
  number={2},
  pages={26},
  year={2023},
  publisher={MDPI}
}
% \printbibliography

\begin{IEEEbiography}[{\includegraphics[width=1in,height=1.25in,clip,keepaspectratio]{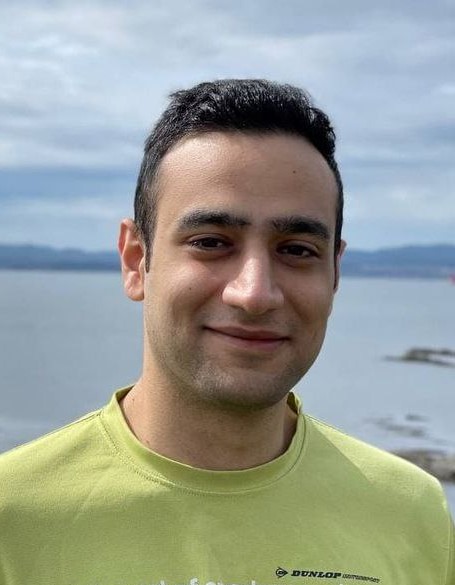}}]{Mehryar Abbasi} s a Ph.D. candidate at Simon Fraser University specializing in deep learning. His research interests include self-supervised learning, time series analysis, and video classification, with a primary application in automated image and video processing. He holds B.Sc. and M.Sc. degrees in Electrical Engineering.
\end{IEEEbiography}
\begin{IEEEbiography}[{\includegraphics[width=1in,height=1.25in,clip,keepaspectratio]{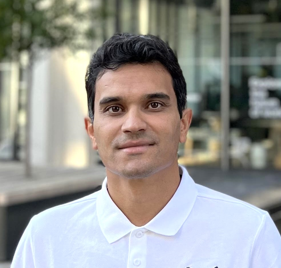}}]{Hadi Hadizadeh} (Member, IEEE) received the M.S. degree in electrical engineering from Iran University of Science and Technology, Tehran, Iran, in 2008, and the Ph.D. degree in engineering science from Simon Fraser University, Burnaby, BC, Canada, in 2013. His current research interests include deep learning, image/video coding and processing, generative AI, and large language models. He was a recipient of the Best Paper Runner-Up Award at ICME 2012 in Melbourne, Australia, and the Microsoft Research and Canon Information Systems Research Australia Student Travel Grant for ICME 2012. In 2013, he was serving as the Vice Chair for the Vancouver Chapter of the IEEE Signal Processing Society.
\end{IEEEbiography}
\begin{IEEEbiography}[{\includegraphics[width=1in,height=1.25in,clip,keepaspectratio]{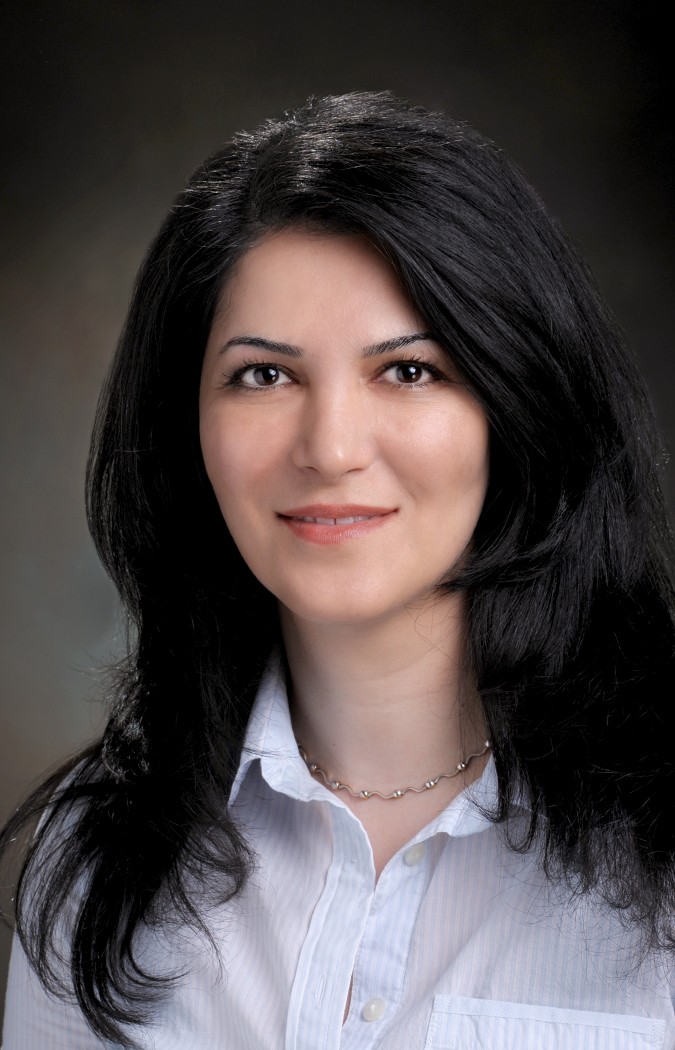}}]{Parvaneh Saeedi} received her M.A.Sc. and Ph.D. degrees from the University of British Columbia in 1998 and 2004,  respectively. From 2004 to 2006, she was with MacDonald, Dettwiler, and Associates Ltd. (MDA) in Richmond, BC, where she worked as a research scientist for various projects in collaboration with Canadian Space Agency, European Space Agency, and Defence Research and Development Canada. Since 2007, she has been with the School of Engineering Science, Simon Fraser University, where she is currently a professor. Her research interests include Medical Image Processing and Analysis, Geospatial Data Processing and Analysis (optical and radar data), 3D Reconstruction, Computer Vision, and Machine Learning.

\end{IEEEbiography}
% \newpage

 

% \end{IEEEbiography}



\vfill

\end{document}